\documentclass[pra,aps,superbib,citeautoscript,twocolumn]{revtex4-1}

\usepackage[colorlinks=true,urlcolor=blue,citecolor=blue,linkcolor=blue]{hyperref}

\usepackage[T1]{fontenc}
\usepackage[utf8]{inputenc} 
\usepackage[english]{babel}
\usepackage{graphicx}
\usepackage{float}
\usepackage{caption}
\usepackage{subcaption}
\usepackage{amsmath, amssymb, amsfonts}
\usepackage{mathtools}
\usepackage{bm}
\usepackage{bbm} 
\usepackage{chemformula}
\usepackage{physics}
\usepackage{tabularx, booktabs}
\usepackage{xspace}
\usepackage{listings}
\usepackage{tikz}
\usetikzlibrary{arrows.meta, shapes}
\usepackage{gnuplottex}
\usepackage{placeins}
\usepackage{caption}
\usepackage[colorlinks=true, linkcolor=blue, citecolor=blue]{hyperref}
\usepackage{algorithm}
\usepackage[noend]{algpseudocode}
\usepackage{algorithm}
\usepackage{algpseudocode}

\definecolor{dgreen}{rgb}{0,.5,0}
\definecolor{dred}{rgb}{.7,.0,.0}

\DeclareMathOperator*{\argmin}{arg\,min}
\DeclareMathAlphabet\mathbfcal{OMS}{cmsy}{b}{n}

\newcommand{\ie}{{\it i.e.}}
\newcommand{\myket}[1]{\left\vert #1\right\rangle}
\newcommand{\mybra}[1]{\left\langle #1\right\vert}
\newcommand{\inner}[2]{\left\langle #1 \middle\vert #2 \right\rangle}
\newcommand{\innerop}[3]{\left\langle #1 \middle\vert #2 \middle\vert #3 \right\rangle}

\newcommand{\boldv}{\bf{v}}

\DeclareMathAlphabet\mathbfcal{OMS}{cmsy}{b}{n}

\newcommand{\undern}{\underline{n}}
\newcommand{\be}{\begin{eqnarray}}
\newcommand{\ee}{\end{eqnarray}}

\begin{document}

\title{
Generalized local potential functional embedding theory of localized orbitals}

\author{Wafa Makhlouf$^a$}
\email{wmakhlouf@unistra.fr}
\author{Bruno Senjean$^b$}
\author{
Emmanuel Fromager$^{a,c}$
}
\affiliation{\it 
~\\
$^a$Laboratoire de Chimie Quantique,
Institut de Chimie, CNRS / Universit\'{e} de Strasbourg,
4 rue Blaise Pascal, 67000 Strasbourg, France\\
}
\affiliation{
$^b$
ICGM, Univ Montpellier, CNRS, ENSCM, Montpellier, France\\
}
\affiliation{$^c$University of Strasbourg Institute for Advanced Study,
5, all\'{e}e du G\'{e}n\'{e}ral Rouvillois, F-67083 Strasbourg, France}


\begin{abstract}

In this work we introduce a generalized flavor, in the sense of generalized Kohn-Sham density functional theory (gKS-DFT), of the recently derived local potential functional embedding theory (LPFET) [J. Chem. Theory Comput. 2025, 21, 20, 10293], where the in-principle exact formalism of DFT is combined with that of density matrix embedding theory (DMET). In generalized LPFET (gLPFET), the embedding clusters are designed from a full-size gKS system where the (in-principle non-local) Hartree-Fock exchange potential is combined with a local (in the localized orbital representation) correlation potential. The latter is optimized self-consistently such that gKS and local embedding cluster's densities match. Unlike in DMET, which uses the same (global) chemical potential value in all clusters, each embedded orbital has its own chemical potential in gLPFET. We show analytically that, when electron correlation is strongly local, the latter potential becomes a simple functional of the correlation potential. Numerical calculations on model systems confirm the high accuracy of gLPFET in this regime, in contrast to DMET. Moreover, we show that gLPFET completely fixes the flaw of LPFET in weaker correlation regimes, through its appropriate description of the Hartree-exchange potential.      
\end{abstract}

\maketitle



\section{Introduction} \label{sec:Introduction}

The accurate description of strong electron correlations in molecular and extended systems remains a central challenge in electronic structure theory. While wave-function-based methods~\cite{RevModPhys.79.291,Alavi24_Faraday_Correlation} provide a systematically improvable accuracy, their computational cost grows exponentially with system size, limiting their applicability to relatively small systems, in general. Density functional theory (DFT)~\cite{Teale2022_DFT_exchange}, on the other hand, offers a computationally efficient framework by recasting the many-body problem in terms of the electronic density, but its practical success relies on approximate exchange-correlation (xc) density functionals which usually become inadequate in strongly correlated systems. Quantum embedding methods \cite{sun2016quantum,wasserman2020quantum,mejuto2024quantum,Evangelista_2025_Concluding,Verma26_Multireference_Embedding,lacombe2020exactfac,requist2021fset,Jiachen24_Interacting-Bath} have emerged in recent years as a powerful approach to bridge the gap between accuracy and efficiency by partitioning a large system into smaller subsystems that are embedded into a simplified environment (usually referred to as quantum bath). Among these approaches, density matrix embedding theory (DMET)~\cite{knizia2012density,knizia2013density, cances2025analysis,wouters2017five,wouters2016practical,chen2014intermediate,cui2020ground,kawano2020comparative, mukherjee2017simple,
sandhoefer2016density,zheng2016ground,zheng2017stripe,Verma26_Multireference_Embedding,cui2025ab} has attracted significant attention due to its ability to capture strong local correlation effects using compact impurity models that are constructed self-consistently from a full-size mean-field-like reference calculation.\\ 

In this paper, we will focus on a specific flavor of DMET, referred to as density embedding theory (DET)~\cite{bulik2014density,bulik2014electron,fulde2017dealing,plat2020entanglement}, where the self-consistency loop of the embedding exploits the diagonal elements of computed (one-electron reduced) density matrices only, in analogy with lattice Kohn-Sham DFT (KS-DFT). Very recently, the authors have derived an in-principle exact (in the sense of lattice KS-DFT) formulation of DET~\cite{LPFET}, which applies to both model and {\it ab initio} Hamiltonians, thus putting previous connections between KS-DFT and DMET~\cite{bulik2014density,bulik2014electron,fulde2017dealing,plat2020entanglement,mordovina2019self,sekaran2022local} in an even more rigorous and general setting. Its simplification in regimes where electron correlations are strong and local lead to a practical local potential functional embedding theory (LPFET)~\cite{LPFET} where, unlike in DET or DMET, each embedded orbital has its own (so-called impurity chemical) potential. The latter was also shown to be a simple functional of the analog in this context of the local (in the lattice representation) Hartree-xc (Hxc) potential~\cite{LPFET}. Numerical calculations on model systems have confirmed the relevance and accuracy of LPFET in strongly correlated regimes. They have also revealed that, when electron correlation is weaker, LPFET fails dramatically. While this is not completely surprizing, as the impurity chemical potential expression LPFET relies on is justified for strong and local electronic interactions only, a rigorous and practical solution to this problem, which reduces drastically the applicability of the method, is needed. This is the purpose of the present work. As it became clear from Ref.~\cite{LPFET} that involving the Hx potential in the evaluation of the impurity chemical potential was causing the problem, we propose a reformulation of the theory such that only the correlation potential contributes. As shown in the following, this can be achieved rigorously by switching from the conventional KS to a generalized KS (gKS) scheme where the Hx potential becomes the (non-local) Hartree--Fock (HF) one. This leads to a local correlation potential functional embedding theory, that we simply refer to as generalized LPFET (gLPFET). While the use of a non-local Hx potential in the full-size reference calculation is quite common in DMET~\cite{wouters2016practical}, the novelty lies more in the formal connection with gKS-DFT and, most importantly, in its combination with an appropriate impurity chemical potential that is {\it not} global (\ie, it varies from one embedded orbital to another). We confirm numerically on model systems that gLPFET completely fixes the flaw of LPFET while preserving its accuracy in strongly correlated regimes.\\  

The paper is organized as follows. After briefly reviewing D(M)ET in Sec. \ref{sec:brief_review_LPFET} and the lattice formulation of gKS-DFT in Sec.~\ref{sec:gKS-DFT}, we merge the two approaches in Sec.~\ref{sec:gLPFET}. A detailed rationale is derived in Appendix~\ref{appendix:rationale_gLPFET}, for completeness.   
Implementation details are given in Sec. \ref{sec:comput_details} and the results are discussed in Sec. \ref{sec:Results}. Finally, our concluding remarks and perspectives are given in Sec. \ref{sec:conclusions}.

\section{Theory} \label{sec:Theory}

\subsection{Quantum embedding in a localized orbital basis}\label{sec:brief_review_LPFET}

Let us start from the general electronic Hamiltonian expression,
\begin{equation}\label{eq:true_Hamiltonian}
\hat{H}=\hat{h}+\hat{U},
\end{equation}
where the one- and two-electron contributions read as follows in second quantization,
\begin{align}
\hat{h} &= \sum_{ij,\sigma} h_{ij}\hat{c}^\dagger_{i\sigma}\hat{c}_{j\sigma},\\
\label{eq:SQ_elec_int_op}
\hat{U} &= \tfrac{1}{2}\sum_{ijkl,\sigma\sigma'} \langle ij|kl\rangle\hat{c}^\dagger_{i\sigma}\hat{c}^\dagger_{j\sigma'}\hat{c}_{l\sigma'}\hat{c}_{k\sigma},
\end{align}
respectively. In the following, we assume that creation operators $\hat{c}^\dagger_{i\sigma}$ generate (when applied to the vacuum state) orthonormal spin-orbitals $\chi_{i\sigma}$ that are {\it localized}, $\sigma=\lbrace \uparrow,\downarrow \rbrace$ referring to the one-electron spin state. The description of strong local electron correlations through quantum embedding motivates this choice~\cite{wouters2016practical}. What follows is independent of the particular localization scheme and applies equally to {\it ab initio} and lattice Hamiltonians. In lattice models, the index $i$ denotes a site while the one- and two-electron integrals ($h_{ij}$ and $\langle ij|kl\rangle$, respectively) become model parameters. Following DMET~\cite{knizia2012density,wouters2016practical}, each localized orbital $\chi_i$ (we use the spin-restricted formalism throughout the paper) can be embedded individually into a so-called quantum bath, consisting here of a single (delocalized over the rest of the full system) bath orbital $b^{(i)}$. Once it has been embedded, $\chi_i$ is usually referred to as {\it impurity}, as the electronic repulsion on that orbital becomes central when dealing with strong local correlations. The bath orbital can be expressed explicitly as follows, in terms of (full-size) one-electron reduced density matrix (1RDM) elements $\gamma^\sigma_{ij}=\langle\hat{c}^\dagger_{i\sigma}\hat{c}_{j\sigma}\rangle$~\cite{sekaran2023unified},
\be\label{eq:bath_orb_simple_exp}
b^{(i)}=\dfrac{1}{\sqrt{\sum_{j\neq i}\left(\gamma^\sigma_{ij}\right)^2}}\sum_{j\neq i}\gamma^\sigma_{ij}\chi_j.
\ee
The reference 1RDM from which the bath is constructed is usually that of a single Slater determinant, thus allowing for insightful connections with KS-DFT, with practical implications~\cite{LPFET}, as further discussed in the following.
Even though it is not considered in the present work, for the sake of simplicity and clarity, multiple orbitals can be embedded simultaneously. In this case, the number of bath orbitals is usually equal to the number of embedded orbitals~\cite{sekaran2023unified}.\\

We now turn to the so-called interacting bath formulation of the embedding theory, which is commonly used in quantum chemistry applications~\cite{wouters2016practical}. In this formulation, the full Hamiltonian is projected onto the ``impurity+bath'' (so-called {\it embedding cluster}) active space as follows,
\begin{align}
\label{eq:bare_projection_onto_the_cluster}
\hat{H}&\rightarrow\hat{P}^{(i)}\hat{H}\hat{P}^{(i)},\\
\label{eq:projector_op_cluster}
\hat{P}^{(i)}&=\sum_\alpha \vert \Phi^{(i)}_{\rm core}\Phi^{(i)}_\alpha\rangle\langle\Phi^{(i)}_{\rm core}\Phi^{(i)}_\alpha\vert,
\end{align}
where $\{\Phi^{(i)}_\alpha\}$ is the collection of Slater determinants obtained by distributing two electrons (in the present single-orbital embedding case~\cite{sekaran2023unified}) among the (two) impurity and bath orbitals, which corresponds to a half-filled embedding cluster. The latter property still holds in the multiple-orbital embedding case~\cite{sekaran2023unified}. The core electrons in Eq.~(\ref{eq:projector_op_cluster}) occupy the orbitals of the reference full-size Slater determinant that have zero overlap with the embedded orbital $\chi_i$. Note that both bath and core orthonormal orbitals can be generated straightforwardly by applying a unitary Householder transformation to the reference (idempotent in the present case) full-size 1RDM~\cite{sekaran2023unified}.\\

As the embedding procedure is guaranteed to be exact at the non-interacting or HF levels of calculation only~\cite{sekaran2021householder,sekaran2023unified,cances2025analysis}, the projected many-electron Hamiltonian of Eq.~(\ref{eq:bare_projection_onto_the_cluster}) is usually complemented by a global (\ie, independent on the orbital $\chi_i$ that is embedded) chemical potential correction $-\mu_{\rm glob}\hat{n}_i$, with $\hat{n}_i=\sum_{\sigma}\hat{c}^\dagger_{i\sigma}\hat{c}_{i\sigma}$, such that the embedding clusters (altogether) reproduce the correct total number $N$ of electrons in the full system. This leads to the DMET (which is equivalent to DET in the present single-orbital embedding case) definition of the embedding cluster's Hamiltonian,
\begin{equation}
\label{eq:DET_cluster_Hamil}
\hat{\mathcal{H}}^{(i)}\overset{\rm DET}{\equiv} \hat{P}^{(i)}\hat{H}\hat{P}^{(i)}-\mu_{\rm glob}\hat{n}_i,
\end{equation}
and the associated electron number constraint
\be
\sum_i\langle\hat{n}_i\rangle_{\hat{\mathcal{H}}^{(i)}}\overset{!}{=}N.
\ee
In a recent paper~\cite{LPFET}, the authors have shown that the embedding procedure can be exactified, in the sense of (lattice) DFT, when the reference full-size Slater determinant is chosen to be the ground state of the following non-interacting KS-like Hamiltonian,
\be
\hat{h}^{\rm KS}=\hat{h}+\sum_i v_i^{\rm Hxc}\hat{n}_i,
\ee
where the analog of the DFT Hxc potential (which, in this context, is fully {\it local} in the localized orbital representation) ensures that $\hat{h}^{\rm KS}$ reproduces the exact ground-state localized orbitals occupation, \ie, 
\be\label{eq:exact_KS_mapping}
\langle\hat{n}_i\rangle_{\hat{h}^{\rm KS}}=\langle\hat{n}_i\rangle_{\hat{H}},\,\forall i.
\ee
These occupations will be referred to as density (or density profile) in the following. Most importantly, an exact density-functional relation has been established in Ref.~\citenum{LPFET} between the local Hxc potential and the embedding chemical potential $\mu_{\rm glob}$ of Eq.~(\ref{eq:DET_cluster_Hamil}), thus showing, in particular, that it is impurity-dependent: $\mu_{\rm glob}\rightarrow \mu^{(i)}_{\rm imp}$. In addition, it was shown that, when electron correlation is strong and local, it can be expressed explicitly as follows,
\begin{equation}\label{eq:original_ind_LPFET_chem_pot}
\mu^{(i)}_{\rm imp}=\sum_{k}\langle b^{(i)} \vert \chi_k \rangle^2 v^{\rm Hxc}_k.
\end{equation}
These findings lead to the LPFET approach~\cite{LPFET}, in which the embedding cluster's Hamiltonian now reads
\be\label{eq:LPFET_emb_cluster_Hamil}
\hat{\mathcal{H}}^{(i)}\overset{\rm LPFET}{\equiv} \hat{P}^{(i)}\hat{H}\hat{P}^{(i)}-\mu^{(i)}_{\rm imp}\hat{n}_i.
\ee
The local Hxc potential, from which $\mu^{(i)}_{\rm imp}$ is evaluated, is the basic variable of the theory (hence the name LPFET). It is determined self-consistently from the following local density constraints,
\be\label{eq:dens_constraint_DET_and_LPFET}
\langle\hat{n}_i\rangle_{\hat{h}^{\rm KS}}\overset{!}{=}\langle\hat{n}_i\rangle_{\hat{\mathcal{H}}^{(i)}},\,\forall i,
\ee
which is an approximation to the exact mapping constraint of Eq.~(\ref{eq:exact_KS_mapping}). Note that both the projector $\hat{P}^{(i)}$ and the bath orbital $b^{(i)}$ are implicit functionals of the Hxc potential, since they are determined from the ground state of $\hat{h}^{\rm KS}$. Let us also stress that DET is using the exact same density constraint as LPFET (the one in Eq.~(\ref{eq:dens_constraint_DET_and_LPFET})). The only difference is that, in DET, the Hxc potential is determined up to a constant shift, in addition to the global chemical potential, while in LPFET, the Hxc potential is uniquely determined from Eq.~(\ref{eq:original_ind_LPFET_chem_pot}).\\

At convergence, the ground-state wavefunctions of the different embedding clusters can be used to compute one- and two-electron RDMs locally, thus offering a drastic simplification of the correlated full-size problem. While being successful in the strongly correlated regime, as expected, LPFET was shown to perform poorly when electron correlation is weaker~\cite{LPFET}. This is not completely surprizing as the relation between the impurity chemical potential and the Hxc potential it relies on (see Eq.~(\ref{eq:original_ind_LPFET_chem_pot})) has been derived away from the latter correlation regime. In the rest of the paper we propose to fix this flaw by optimizing the correlation potential only, while using the exact (HF-like) Hx potential, which is in principle non-local in the localized orbital basis. Thus, we move from the KS to the gKS formulation of (lattice) DFT, as discussed further in the next section.   

\subsection{Lattice formulation of generalized KS-DFT}\label{sec:gKS-DFT}

By analogy with conventional DFT, which is formulated in the continuous real space, the ground-state energy of the physical second-quantized Hamiltonian (see Eqs.~(\ref{eq:true_Hamiltonian})-(\ref{eq:SQ_elec_int_op})) can be determined, in principle exactly [``exact'' meaning Full Configuration Interaction (FCI) in this context], as a functional of the density profile in the localized orbital basis~\cite{LPFET}. For that purpose, we separate in the one-electron Hamiltonian $\hat{h}$ the off-diagonal terms
\begin{equation}
\hat{T}:=\sum_{i\neq j}h_{ij}\sum_{\sigma}\hat{c}_{i\sigma}^\dagger\hat{c}_{j\sigma},
\end{equation} 
which together play the role of a kinetic energy (hopping) operator in the lattice of localized orbitals, from the diagonal ones,
\begin{equation}
\hat{V}^{\rm ext}:=\sum_i v_i^{\rm ext}\hat{n}_i \quad \text{with } v_i^{\rm ext}=h_{ii},
\end{equation}
which play the role of the local external potential (on-site energy contributions) in the lattice, so that the full Hamiltonian reads, like in DFT,
\begin{equation}\label{eq:decomp_full_Hamilt}
    \hat{H}= \hat{T}+\hat{U}+\hat{V}^{\rm ext}.
\end{equation}
In this context, the Levy--Lieb density functional is defined as follows~\cite{LPFET}, 
\be\label{eq:HK_func_def}
F(\undern)
=\min_{\Psi\rightarrow \undern}\innerop{\Psi}{\hat{T}+\hat{U}}{\Psi},
\ee
for any $N$-representable density profile 
\be
\undern\equiv\left\{n_i:=\langle\hat{n}_i\rangle\right\},\; \sum_i n_i=N,
\ee
and the exact ground-state energy is obtained from the following variational principle,
\be
E=\min_{\underline{n}}\left\{F(\underline{n})+\sum_iv^{\rm ext}_in_i\right\}.
\ee
Note that, in Eq.~(\ref{eq:HK_func_def}), the $N$-electron density constraint $\Psi\rightarrow \undern$ reads more explicitly as follows,
\be
\undern^{\Psi}\equiv\left\{n_i^{\Psi}:=\langle\Psi \vert\hat{n}_i\vert \Psi\rangle\right\}=\undern, 
\ee
$\Psi$ being {\it any} (\ie, not necessarily a Slater determinant) trial $N$-electron wavefunction.\\ 

While in conventional (KS) DFT and LPFET, the Levy--Lieb functional is split into its non-interacting (kinetic energy) analog and the remaining Hxc energy part, we follow a different path in this work, motivated by our wish to improve the description of the Hx potential in LPFET. We consider the following partitioning instead,
\begin{equation}\label{eq:Hxc_ener_exp_from_F}
F(\underline{n}) = F_{\rm HF}(\underline{n}) + E_{\rm c}(\underline{n}),
\end{equation}
where the HF version of Levy--Lieb's functional, 
\begin{equation}\label{eq:def_F_HF_func}
F_{\rm HF}(\underline{n}) = \min_{\Phi \to \underline{n}} \langle \Phi \vert \hat{T} + \hat{U}\vert \Phi \rangle, 
\end{equation}
is defined through a density constrained search over {\it single} $N$-electron Slater determinants $\Phi$ only, 
\ie,
\be\label{eq:dens_constraint_Slater_det}
\undern^{\Phi}\equiv\left\{n_i^{\Phi}:=\langle\Phi \vert\hat{n}_i\vert \Phi\rangle\right\}=\undern, 
\ee
while the missing correlation energy is described with the appropriate correlation density functional (we use the same notation $E_{\rm c}(\underline{n})$ as in regular KS-DFT or LPFET but the functional is different, strictly speaking),
\begin{equation}\label{eq:corr_func_gKS-DFT}
E_{\rm c}(\underline{n}) = \innerop{\Psi(\undern)}{\hat{T}+\hat{U}}{\Psi(\undern)}  - \innerop{\Phi(\undern)}{\hat{T}+\hat{U}}{\Phi(\undern)},  
\end{equation}
$\Psi(\undern)$ and $\Phi(\undern)$ being the density-functional minimizers in Eqs.~(\ref{eq:HK_func_def}) and (\ref{eq:def_F_HF_func}), respectively.  
Note that we use 100$\%$ of HF exchange in Eq.~(\ref{eq:def_F_HF_func}), while, in practical (real space) DFT, a fraction of it, combined with a complementary density-functional exchange energy is preferred~\cite{Teale2022_DFT_exchange}. In the present context, both exchange and correlation density-functional energies are ultimately evaluated as orbital-dependent quantities, through the embedding of the localized orbitals. In other words, we never use (approximate) functionals that depend explicitly on the density (\ie, the localized orbitals occupation in this context), hence the separation adopted in Eqs.~(\ref{eq:Hxc_ener_exp_from_F}) and (\ref{eq:def_F_HF_func}).\\

From the exact decomposition of Eq.~(\ref{eq:Hxc_ener_exp_from_F}), the ground-state energy can be expressed as follows (see Eq.~(\ref{eq:decomp_full_Hamilt}) and the density constraint of Eq.~(\ref{eq:dens_constraint_Slater_det})), 
\be
E=\min_{\underline{n}}\left\{
\min_{\Phi \to \underline{n}}\left\{ \langle \Phi \vert \hat{H}\vert \Phi \rangle
+E_{\rm c}(\underline{n}^{\Phi})\right\}\right\},
\ee
thus leading to the final energy expression
\be\label{eq:VP_gKS-DFT}
E=\min_{\Phi\rightarrow N}\left\{\langle\Phi \vert\hat{H}\vert \Phi \rangle+E_{\rm c}(\underline{n}^\Phi)\right\},
\ee
where we recall that the minimization is restricted to single Slater determinants. Unlike in a regular HF calculation, the exact ground-state energy is in principle recovered, thanks to the complementary correlation density functional. The minimizing Slater determinant in Eq.~(\ref{eq:VP_gKS-DFT}) is the ground state of the following {\it generalized} KS (gKS) Hamiltonian~\cite{seidl1996generalized},
\begin{equation}\label{eq:sc_gKS_equation}
    \hat{h}^{\rm gKS} = \hat{T} + \hat{V}^{\rm ext} + \hat{V}^{\rm Hx} + \sum_i v_i^{\rm c}\hat{n}_i,
\end{equation}
and, by construction, it reproduces the exact ground-state density profile, \ie, that of the $N$-electron ground state $\Psi_0$ of $\hat{H}$: 
\be\label{eq:exact_dens_mapping_gKS}
\left\{\langle\hat{n}_i\rangle_{\hat{h}^{\rm gKS}}\right\}=\left\{\langle\hat{n}_i\rangle_{\hat{H}}\right\}=:\underline{n}^{\Psi_0}.
\ee
Note that the (HF-like) Hx potential operator  
\be
\hat{V}^{\rm Hx}=\sum_{ij\sigma} v^{\rm Hx}_{ij\sigma}\hat{c}_{i\sigma}^\dagger\hat{c}_{j\sigma},
\ee
with (for a closed-shell system)
\be\label{eq:non-local_HF_Hx_pot}
v^{\rm Hx}_{ij\sigma}=v^{\rm Hx}_{ij}=\sum_{kl} \left( 2\langle ik | jl \rangle -\langle ik | lj \rangle \right) \gamma^\sigma_{kl},
\ee
is not exclusively expressed in terms of the density, unlike in KS-DFT (hence the name ``generalized''~\cite{seidl1996generalized}). In the present case, the (gKS) 1RDM elements $\gamma^\sigma_{kl}\equiv \langle\hat{c}^\dagger_{k\sigma}\hat{c}_{l\sigma}\rangle_{\hat{h}^{\rm gKS}}$ are needed for its evaluation, which must be achieved self-consistently, like in a regular HF calculation. Note also that, as readily seen from Eq.~(\ref{eq:non-local_HF_Hx_pot}), the Hx potential of gKS-DFT will be {\it non-local} in the lattice representation, in general, unlike in lattice KS-DFT. On the other hand, the {\it local} density-functional correlation potential contribution to the gKS Hamiltonian, which can be formally expressed as follows (by allowing more general electron number non-conserving density variations for an appropriately fixed chemical potential value), 
\be\label{eq:exact_corr_pot_gKS}
v_i^{\rm c}\equiv \left.\dfrac{\partial E_{\rm c}({\bf n})}{\partial n_i}\right|_{{\bf n}=\{\langle\hat{n}_i\rangle_{\hat{h}^{\rm gKS}}\}}, 
\ee
 brings a second ``layer'' of self-consistency. In the above expression, ${\bf n}\equiv \{n_i\}$ denotes a more general density profile than the $N$-representable one $\underline{n}$, as it does not necessarily integrate to the integer number $N$ of electrons in the system under study. The fact that each localized orbital occupation $n_i$ in $\bf n$ can be treated as an independent (from the other occupations) variable in the functional derivative of Eq.~(\ref{eq:exact_corr_pot_gKS}) will be central later (in Appendix~\ref{appendix:rationale_gLPFET}) for adapting the LPFET impurity chemical potential expression of Eq.~(\ref{eq:original_ind_LPFET_chem_pot}) to a gKS reference calculation. According to Ref.~\citenum{LPFET}, it should be possible to derive a formally exact (lattice) density-functional embedding theory based on the gKS scheme by using electron number conserving density variations only. This is left for future work.        

\subsection{Local correlation potential functional embedding theory}\label{sec:gLPFET}

While the LPFET briefly sketched in Sec.~\ref{sec:brief_review_LPFET} relies on a reference lattice-like KS-DFT calculation, in which the full Hxc potential is optimized self-consistently (as a whole) {\it via} the impurity chemical potential expression of Eq.~(\ref{eq:original_ind_LPFET_chem_pot}), our desired generalized flavor of LPFET should rely on the gKS-DFT of Sec.~\ref{sec:gKS-DFT} instead. This immediately raises the question of what the expression of Eq.~(\ref{eq:original_ind_LPFET_chem_pot}) should be in this case. 
There is no obvious nor clear answer to that question since, in gKS-DFT, there is no local Hx potential, unlike in the formula of Eq.~(\ref{eq:original_ind_LPFET_chem_pot}). In the following, we investigate and simplify the latter further in the strongly correlated regime, where it finds its justification, before deriving (in Appendix~\ref{appendix:rationale_gLPFET}) a more detailed and rigorous rationale.\\  


Let us consider the embedding of a specific localized orbital $\chi_i$ and assume that electronic repulsions are essentially local and strong, so that, from the perspective of $\chi_i$, the interaction operator (see Eq.~(\ref{eq:SQ_elec_int_op})) can be simplified drastically as follows~\cite{LPFET},
\be
\hat{U}\approx \langle ii\vert ii\rangle\,\hat{n}_{i\uparrow}\hat{n}_{i\downarrow},
\ee
where $\hat{n}_{i\sigma}=\hat{c}_{i\sigma}^\dagger\hat{c}_{i\sigma}$. This leads to the following approximate density-functional Hx energy expression, in the context of lattice KS-DFT,
\be
E_{\rm Hx}({\bf n})\approx  \langle ii\vert ii\rangle\, n_{i\uparrow}n_{i\downarrow}=\dfrac{1}{4}\langle ii\vert ii\rangle\, n_i^2,
\ee
and the corresponding (local) Hx potential, 
\be
v_j^{\rm Hx}\underset{n_i=n^{\Psi_0}_i}{\approx}\dfrac{\delta_{ij}}{2}\langle ii\vert ii\rangle\, n_i,  
\ee
which becomes a functional of the {\it local} density $n_i$. As it vanishes everywhere except on the embedded orbital $\chi_i$, it does {\it not} contribute to the summation in the right-hand side of Eq.~(\ref{eq:original_ind_LPFET_chem_pot}), since the bath orbital $b^{(i)}$ is orthogonal to $\chi_i$, by construction (see Eq.~(\ref{eq:bath_orb_simple_exp})). As a result, the impurity chemical potential can be expressed solely in terms of the correlation potential:
\begin{equation}\label{eq:ind_LPFET_chem_pot}
\mu^{(i)}_{\rm imp}\approx \sum_{k}\langle b^{(i)} | \chi_k \rangle^2 v^{\rm c}_k.
\end{equation}
Note that, so far, we have derived the above approximation (which is perfectly justified in the strongly correlated regime) in the context of regular LPFET. In the latter case, the embedding cluster is constructed from a full-size (non-interacting) KS system where the Hxc potential is {\it local} in the localized orbital representation. We show in Appendix~\ref{appendix:rationale_gLPFET} that Eq.~(\ref{eq:ind_LPFET_chem_pot}) still holds if the embedding cluster is constructed from the gKS system instead. This is the key formal result of this work.\\

Combining the impurity chemical potential ansatz of Eq.~(\ref{eq:ind_LPFET_chem_pot}) with the following local density constraints (see Eq.~(\ref{eq:LPFET_emb_cluster_Hamil}), where the embedding cluster's Hamiltonian $\hat{\mathcal{H}}^{(i)}$ is defined),
\be
\langle\hat{n}_i\rangle_{\hat{h}^{\rm gKS}}\overset{!}{=}\langle\hat{n}_i\rangle_{\hat{\mathcal{H}}^{(i)}}, \forall i,
\ee
which approximates the exact density mapping of Eq.~(\ref{eq:exact_dens_mapping_gKS}), defines a local correlation potential functional embedding theory that we simply refer to as gLPFET in the rest of the paper. Note that, in gLPFET, the bath orbitals $b^{(i)}$ (and, therefore, the projectors $\hat{P}^{(i)}$) are implicit functionals of the to-be-determined correlation potential ${\bf v}^{\rm c}\equiv \{v_k^{\rm c}\}$. 

\section{Implementation and computational details}\label{sec:comput_details}

We apply the gLPFET approach introduced in Sec.~\ref{sec:gLPFET} to two benchmark systems for which exact FCI solutions can be computed for comparison. The first one is a non-uniform finite Hubbard ring consisting of $L=6$ lattice sites at half-filling. We impose periodic boundary conditions and set the nearest-neighbor hopping amplitude to unity ($t=1$). The Hamiltonian is written as
\begin{equation}\label{eq:Hubbard_Hamil}
    \begin{split}
    \hat{H}&=-t\sum^{L-1}_{i=0}\sum_{\sigma}\left(\hat{c}_{i\sigma}^\dagger\hat{c}_{i+1\sigma}+\hat{c}_{i+1\sigma}^\dagger\hat{c}_{i\sigma}\right)
    \\
    &\quad
    +U\sum^{L-1}_{i=0}\hat{n}_{i\uparrow}\hat{n}_{i\downarrow}
    +\sum^{L-1}_{i=0}v^{\rm ext}_i\hat{n}_i
    ,
    \end{split}
\end{equation}
where the external on-site potential introduces the non-uniformity and is chosen as $\left\{v^{\rm ext}_i\right\}_{0\leq i\leq 5}\equiv \left\{-1, 2, -2, 3, -3, 1\right\}$. The second benchmark system is an \emph{ab initio} linear chain of six equispaced hydrogen atoms (H$_6$) in the minimal STO-3G basis. To obtain a localized basis suitable for the local embedding formalism, the atomic orbitals are symmetrically orthogonalized using Löwdin’s method~\cite{lowdin1950non}, and these Orthogonalized Atomic Orbitals (OAOs) are used for all subsequent calculations.
The implementation follows the theoretical derivation in Sec.~\ref{sec:Theory}, specifically utilizing the separation between the non-local (HF-like) Hx potential from the local correlation potential. The numerical simulations were performed using the QuantNBody package~\cite{yalouz2022quantnbody} (developed by S. Yalouz et al.), which facilitated the manipulation of the quantum many-body operators required for our scheme.
\begin{enumerate}
    \item  \textbf{Inner Loop: gKS-SCF reference}. The gKS Hamiltonian is constructed as: 
    \begin{equation} 
    \hat{h}^{\rm gKS} = \hat{h} + \sum_i v_i^{\rm c} \hat{n}_i + \hat{v}^{\rm Hx}[\boldsymbol{\gamma}] \end{equation} where $\hat{v}^{\rm Hx}[\boldsymbol{\gamma}] $ is the non-local HF potential operator (represented as $2\hat{J}-\hat{K}$ in the restricted closed-shell formalism) constructed from the density matrix (1RDM) $\bm\gamma$. The density matrix is updated by diagonalizing $\hat{h}^{\rm gKS}$ and occupying the lowest $N_{occ}$ orbitals. We employ the Direct Inversion in the Iterative Subspace (DIIS) to accelerate convergence utilizing the standard routine available in the open-source Psi4 package~\cite{smith2020psi4}. 
    \item \textbf{Embedding Cluster Construction}: 
    Once the inner gKS-SCF converges, we compute the gKS density profile $\underline{n}^{\rm gKS} = \{\langle \hat{n}_i \rangle_{\hat{h}^{\rm gKS}}\}$. For each  site $i$, we construct the embedding cluster by applying the Householder transformation~\cite{sekaran2021householder,yalouz2022quantum,sekaran2023unified} to the converged gKS density matrix $\bm\gamma$. Crucially, as derived in Eq.~(\ref{eq:ind_LPFET_chem_pot}), the impurity chemical potential for the corresponding cluster depends only on the local correlation potential $\mathbf{v}^{\rm c}$.
    
    \item \textbf{Cluster 
    Solution and Outer Optimization}: 
    The (interacting) embedding cluster's Hamiltonian is solved using exact diagonalization to obtain the correlated local density $n_i^{\rm cl}$. The deviation between the collection of cluster densities and the reference full-size gKS density profile serves as the objective function for the outer loop:
    \begin{equation}\label{eq:error_density}
    \Delta(\mathbf{v}^{\rm c}) = \sqrt{\sum_i \left( n_i^{\rm cl}(\mathbf{v}^{\rm c}) - n_i^{\rm gKS}(\mathbf{v}^{\rm c})\right)^2}
    \end{equation}
    A standard optimizer updates the local potential vector $\mathbf{v}^{\rm c}$ to minimize this error. Then fed back into the gKS Hamiltonian (Step 1), and the entire cycle repeats. This procedure continues until the density mapping is achieved, which yields the correlation potential that best reproduces the correlated density profile of the clusters within the gKS framework.
\end{enumerate}

The gLPFET algorithm chart reads as follows.

\begin{algorithm}[H]
\caption{gLPFET embedding loop}
\begin{algorithmic}[1]
\State \textbf{Input:} Hamiltonian integrals $h_{ij}$, $\langle ij\vert kl\rangle$, electron number $N_{el}$
\State \textbf{Output:} Converged local correlation potential $\boldv^{c}$, HF-like potential $\boldv^{\rm Hx}$, final densities

\State Initialize $\boldv^{c}\gets\mathbf{0}$, $\boldv^{\rm Hx}\gets\mathbf{0}$, \State \textbf{Outer Loop: Correlation Potential Optimization}

\While{not converged}
\State \textbf{Inner Loop: gKS-SCF (Update Non-Local Exchange)}
\State Initialize density matrix $\boldsymbol{\gamma}$
\While{SCF not converged}
\State Build non-local HF potential: $\mathbf{v}^{\rm Hx} = 2\mathbf{J}[\boldsymbol{\gamma}] - \mathbf{K}[\boldsymbol{\gamma}]$
\State Build gKS Hamiltonian: $\mathbf{h}^{\rm gKS} =  \mathbf{h} + \mathbf{v}^{\rm Hx} + \mathbf{v}^{\rm c}$
\State Diagonalize $\mathbf{h}^{\rm gKS} \to$ New orbitals $\to$ Update $\boldsymbol{\gamma}$
\State Apply DIIS extrapolation
\EndWhile
\State Compute gKS densities: $n_i^{\rm gKS} = \gamma_{ii}$
\State \textbf{Embedding Step }
\For{each site $i$}
\State Perform Householder transformation on $\boldsymbol{\gamma}$
\State Map impurity chem. potential: $\mu_{\rm imp}^{(i)} \gets \text{Map}(\mathbf{v}^{\rm c})$
\State Solve cluster Hamiltonian $\hat{\mathcal{H}}^{(i)}$
\State Extract correlated site density $n_i^{\rm cl}$
\EndFor
\State \textbf{Update Local Potential}
\State Compute Error $\Delta(\mathbf{v}^{\rm c})$ [Eq.~(\ref{eq:error_density})]
\If{$\Delta (\mathbf{v}^{\rm c}) < \varepsilon_{\rm tol}$}
\State \textbf{break}
\Else
\State Update $\mathbf{v}^{\rm c}$ using standard optimizer
\EndIf
\EndWhile
\State \textbf{Return:} $\boldv^{c}$, $\boldv^{\rm Hx}$, $ \underline{n}^{\rm cl}  $ 

\end{algorithmic}
\end{algorithm}

\section{Results and discussion} \label{sec:Results}

\subsection{Non-uniform Hubbard ring}\label{Hubbard_ring}

\begin{figure}
    \centering
    \includegraphics[width=1\linewidth]{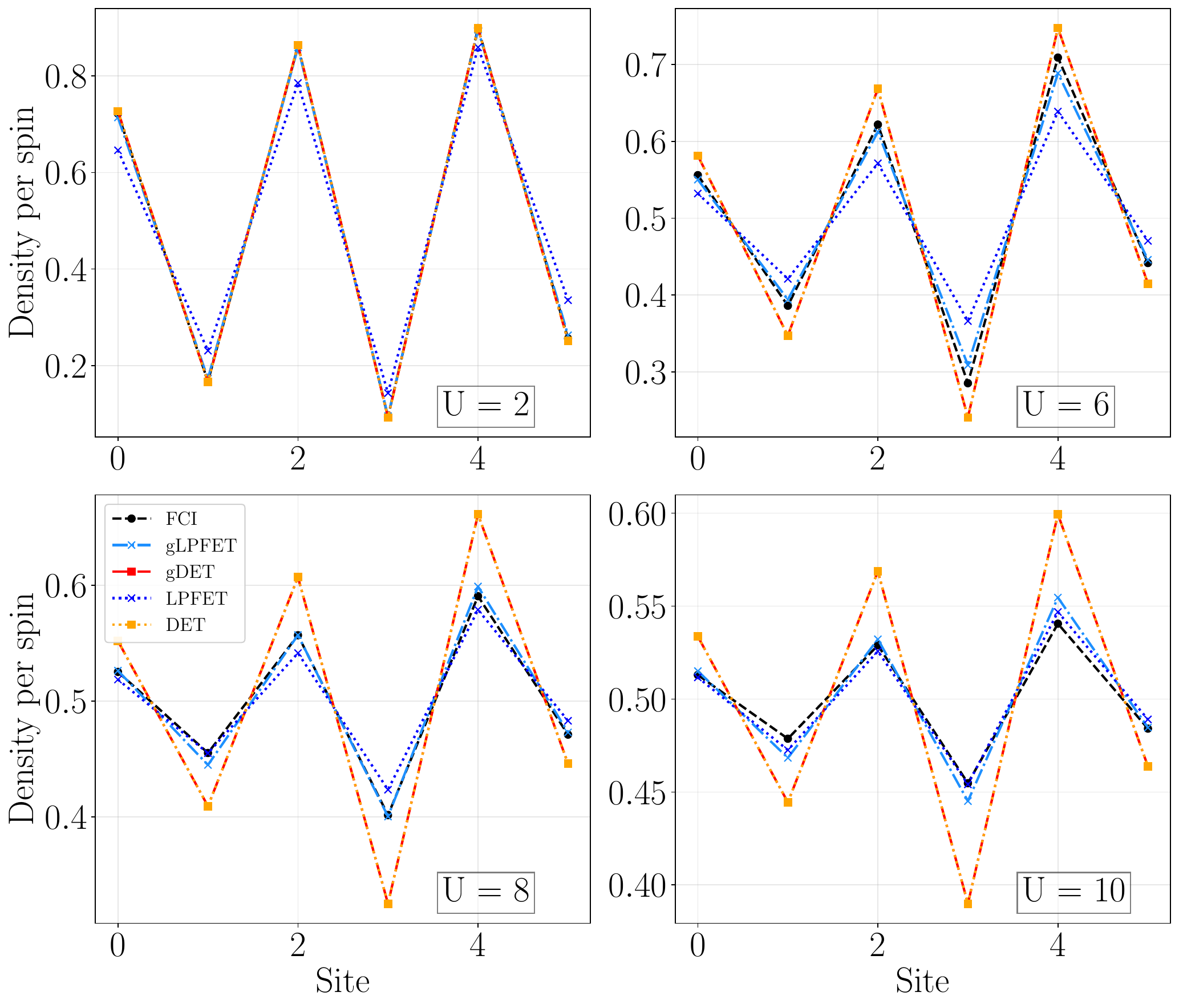}
    \caption{Density profile of the half-filled six-site Hubbard ring obtained for different interaction strengths $U/t$ with (g)LPFET and (g)DET embedding methods. Comparison is made with the exact FCI results.}
    \label{fig:comparison}
\end{figure}

We begin our discussion by comparing in Fig.~\ref{fig:comparison} the self-consistently converged density profiles obtained from the original LPFET~\cite{LPFET} with those computed within the newly proposed gLPFET in the half-filled non-uniform six-site Hubbard ring described in Eq.~(\ref{eq:Hubbard_Hamil}), for various interaction strengths. All results are benchmarked against FCI, which provides the exact solution for this model. Comparison is also made with DET. Note that, for Hubbard Hamiltonians with on-site interactions only, the generalized version of DET (gDET), where a global chemical potential is still used in each embedding cluster~\cite{wouters2016practical}, is equivalent to DET because, unlike in the general {\it ab initio} case, the exchange potential remains local in the lattice representation, according to Eq.~(\ref{eq:non-local_HF_Hx_pot}):
\be
v^{\rm Hx}_{ij}\rightarrow\dfrac{\delta_{ij}}{2}Un_i.
\ee
This equivalence is readily seen from Fig.~\ref{fig:comparison}. The reason why LPFET and gLPFET differ, on the other hand, is that they rely on different impurity chemical potential expressions (see Eqs.~(\ref{eq:original_ind_LPFET_chem_pot}) and (\ref{eq:ind_LPFET_chem_pot}), respectively). As expected and confirmed numerically, in the weakly correlated regime, gLPFET is completely cured from the significant errors that the impurity chemical potential expression regular LPFET relies on (which, we recall, was obtained in the strongly correlated limit) introduces into the self-consistent evaluation of the density. While LPFET is able to capture local features of electron correlation as the interaction strength grows, unlike DET (see the $U/t=6$ and $U/t=8$ panels of Fig.~\ref{fig:comparison}), gLPFET benefits additionally from its improved description of the Hx potential in this regime, thus resulting in highly accurate density profiles. These major improvements over LPFET preserve the accuracy of the method in stronger correlated regimes, as expected. Interestingly, in the intermediate $U/t=10$ case between weakly and strictly correlated regimes, gLPFET and LPFET are not completely identical (they both perform well, LPFET being slightly more accurate), thus indicating a non-negligible impact of the local Hx potential on the impurity chemical potential in this case. 

\begin{figure}
    \centering
    \includegraphics[width=1\linewidth]{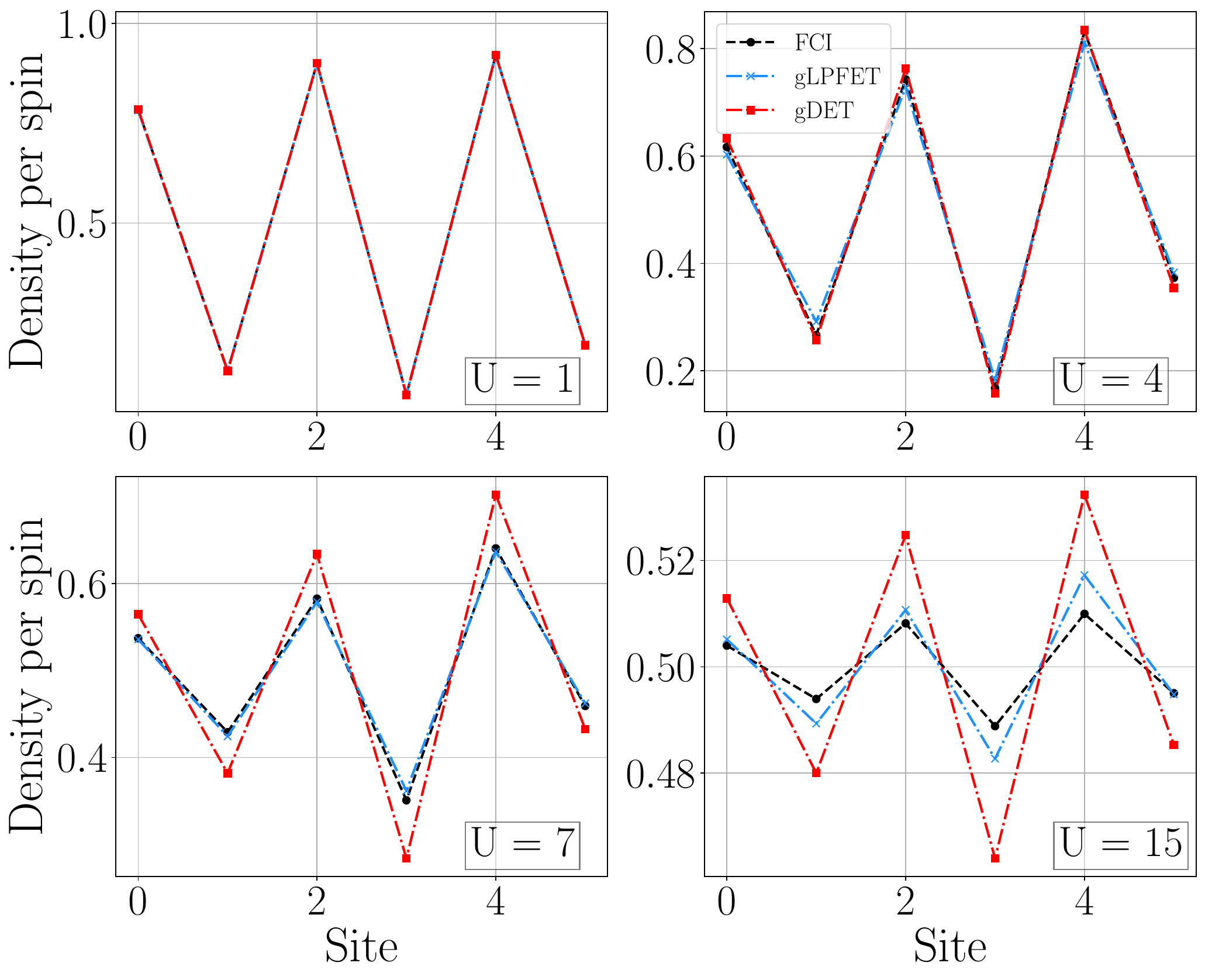}
    \caption{Same as Fig.~\ref{fig:comparison} for generalized embedding methods only and other interaction strengths.}
    \label{fig:Hubbard_1D_strong}
\end{figure}

We now focus on the comparison of gLPFET with the more standard gDET. As expected, when electron correlation is weak (see Fig.~\ref{fig:Hubbard_1D_strong}), both gLPFET and gDET reproduce the FCI densities with very high accuracy. In this regime, the density variation imposed by the external potential dominates over correlation effects, and the correlation potential converges reliably. As $U/t$ increases, correlation effects become more pronounced, and differences in the density profile begin to appear. Up to $U/t=6$, gLPFET continues to perform well, while gDET tends to overestimate or underestimate local densities, reflecting the limitations of its global chemical potential in strongly correlated inhomogeneous systems. For the larger $U/t=8$ and $U/t=10$ interaction strength values (see Fig.~\ref{fig:comparison}), both embedding methods show visible deviations from FCI, but gLPFET remains quantitatively closer across all sites. In such a strongly correlated regime, the density becomes increasingly dominated by on-site repulsions rather than by the external potential, making the density mapping problem more challenging to converge. Nonetheless, the local character of the gLPFET (impurity chemical) embedding potential makes it superior (relative to gDET) in terms of numerical stability (see also Fig.~\ref{fig:Hubbard_1D_strong}).\\  

\begin{figure}
    \centering
    \includegraphics[width=1\linewidth]{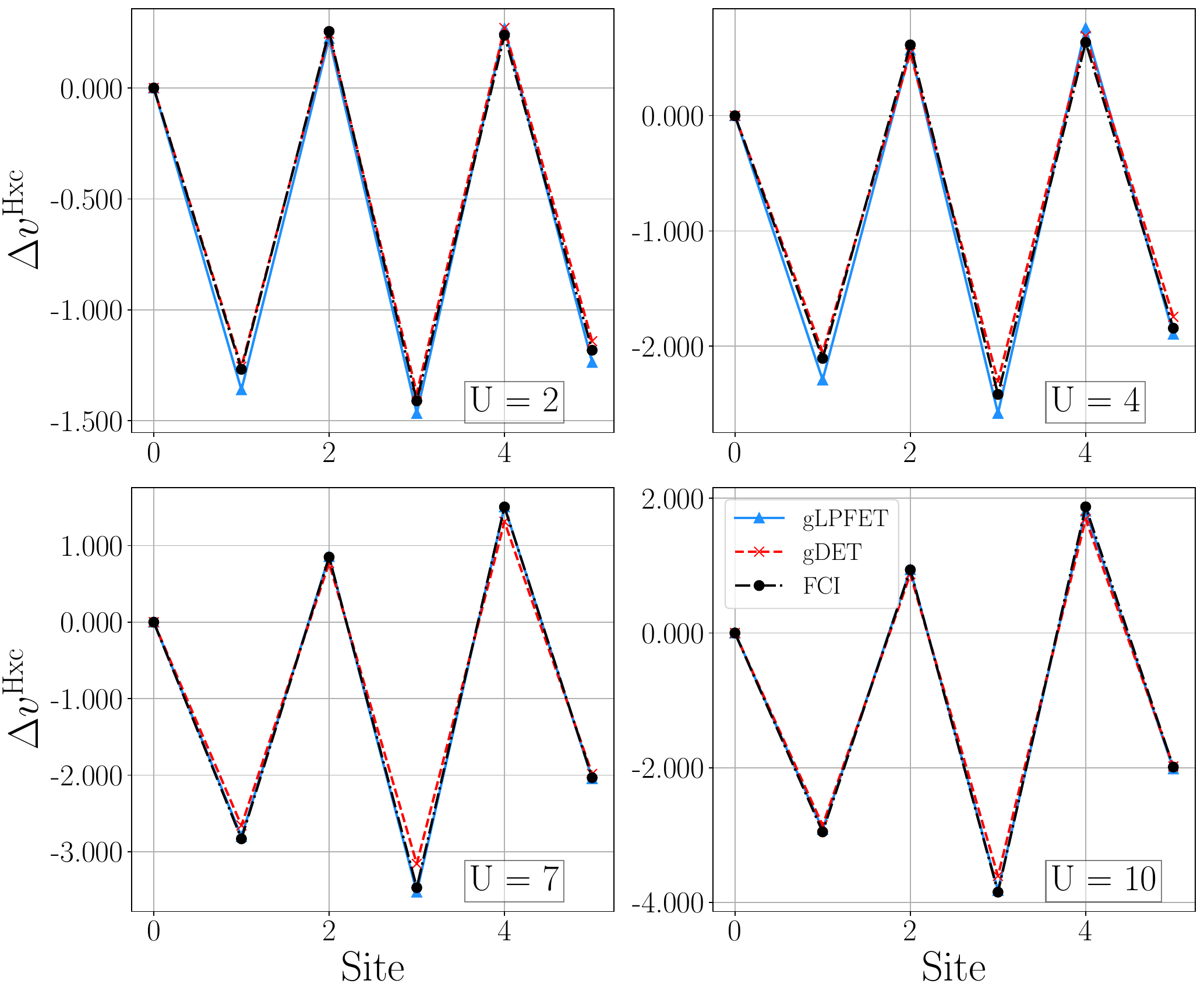}
    \caption{Full Hxc potential on-site values evaluated with respect to that of site 0 at the gLPFET and gDET levels of embedding, for different interaction strengths. Comparison is made with the exact reverse engineered Hxc potential obtained from the FCI density profile.}
    \label{fig:Hxc_potential_weak}
\end{figure}

For such a small model system, which is exactly solvable, it is possible to determine the exact correlation potential, to which we add the (local here) Hx potential, from the FCI density profile. Results are shown and compared in Fig.~\ref{fig:Hxc_potential_weak} with those obtained self-consistently with the generalized embedding methods. As expected, the latter, which are overall relatively accurate, are consistent with the density profiles discussed previously. We note in particular that, for $U/t=4$, the Hxc potential of gLPFET is too attractive on sites 1, 3, and 5, which explains the overestimation of their occupancy (as shown in Fig.~\ref{fig:Hubbard_1D_strong}). On the other hand, in the stronger $U/t=7$ correlation regime, the Hxc potential of gDET is too repulsive on these same sites (that of gLPFET is, on the other hand, highly accurate), which explains the underestimation of local gDET densities in this case (see Fig.~\ref{fig:Hubbard_1D_strong}).\\

\begin{figure}
    \centering
    \includegraphics[width=1\linewidth]{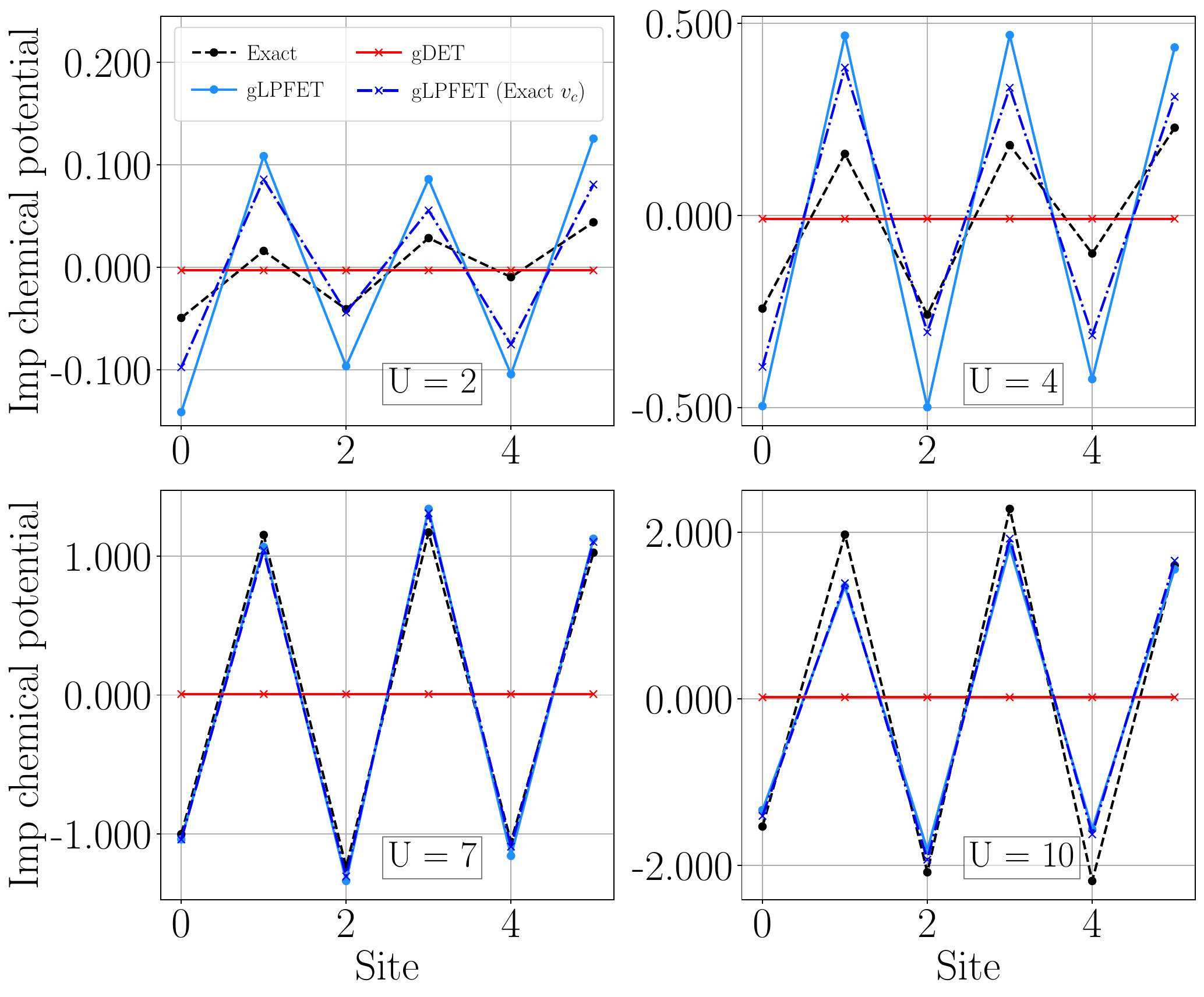}
    \caption{Impurity chemical potential values obtained on each embedded site at the gLPFET and gDET levels of embedding, for various interaction strengths. Comparison is made with the exact reverse engineered values (from the FCI density profile) and those computed from the gLPFET ansatz of Eq.~(\ref{eq:ind_LPFET_chem_pot}) with the exact correlation potential.}
    \label{fig:imp_chem_pots}
\end{figure}

For additional insight, we compare in Fig.~\ref{fig:imp_chem_pots} the exact impurity chemical potential value for each embedded site (which can be determined from the FCI density profile by reverse engineering) with that of gLPFET and gDET. In the weakly correlated $U/t=2$ case, the exact impurity chemical potential profile deviates less than that of gLPFET from the global value of gDET, which is not completely surprizing as the ansatz of Eq.~(\ref{eq:ind_LPFET_chem_pot}) gLPFET relies on has been derived in the strongly correlated regime. This discrepancy is somehow reflected in the Hxc potential shown in Fig.~\ref{fig:Hxc_potential_weak}. For analysis purposes, we implemented the expression of Eq.~(\ref{eq:ind_LPFET_chem_pot}) with the exact correlation potential. As reverse engineering on the gKS system determines the correlation potential up to a constant shift, the remaining global shift was fixed such that the embedding cluster densities reproduce the FCI density profile as closely as possible. For $U/t=2$, we notice that the corresponding impurity chemical potential differs from the exact and gLPFET ones, thus confirming the inaccuracy (which is not dramatic though) of the gLPFET ansatz for weakly correlated systems. On the other hand, as the interaction strength $U/t$ increases, the three methods give very similar results, while gDET is inherently unable to reproduce the variation of the impurity chemical potential from one embedded orbital to another. This nicely validates numerically the rationale of gLPFET derived in Appendix~\ref{appendix:rationale_gLPFET}.
\\

\begin{figure}
    \centering
    \includegraphics[width=1.6\linewidth]{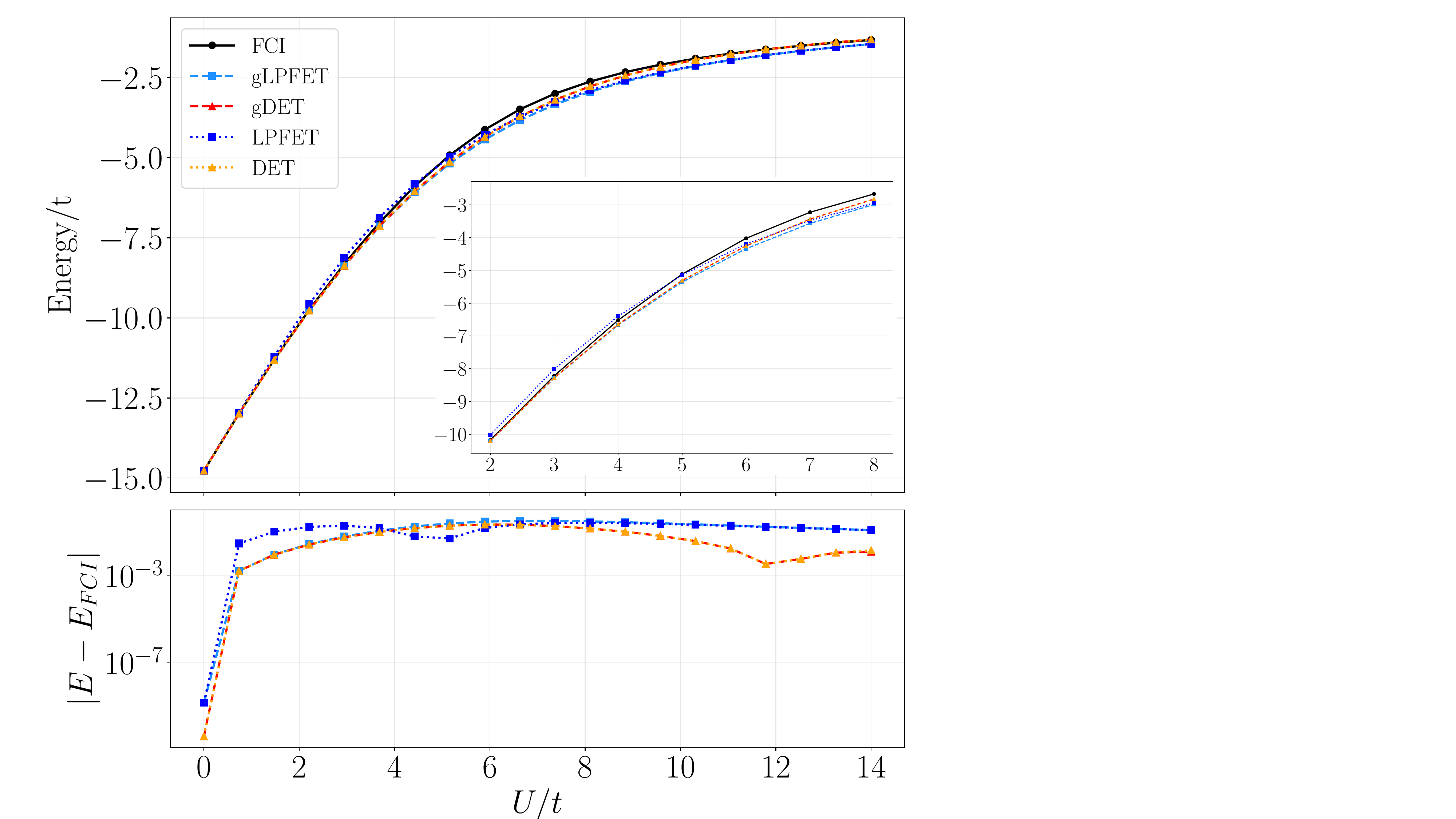}
    \caption{Total energy of the half-filled six-site Hubbard ring computed as a function of the interaction strength $U/t$ with the various embedding methods. Comparison is made with FCI.} 
    \label{fig:Hubb_Energy}
\end{figure}

For completeness, we computed the total energy of the ring using the standard democratic evaluation of 1- and 2RDMs from the correlated wavefunctions of the different embedding clusters (see Refs.~\citenum{wouters2016practical} and \citenum{LPFET} for further details). The energy is plotted as a function of $U/t$ in Fig.~\ref{fig:Hubb_Energy}, where gLPFET is shown to perform much better than the original LPFET in the moderately correlated $2\leq U/t \leq 4$ regime. Therefore, in this case, the drastic improvement in density profile provided by the generalized formulation of the theory is reflected in the total energy, which follows more closely the FCI reference. On the other hand, in the strongly correlated $U/t>8$ regime, the two methods become practically indistinguishable. If we now compare with (g)DET, gLPFET achieves similar accuracy when $U/t<5$. However, in the latter regime, gDET tends to overestimate the correlation energy, yielding values systematically lower than the exact FCI benchmark. As a final note, the better performance of gLPFET (over LPFET) in reproducing the FCI density profile accurately, for $4<U/t<6$, is not reflected at all in the energy plot. The same statement can be made when comparing (g)LPFET to (g)DET, for $U/t\geq 6$. While we suspect error cancellations to come into play, we may question, more generally, the non-variational evaluation of the total energy as well as the democratic evaluation of the RDM elements for that purpose. This can be related to the $N$-representability problem of quantum embedding theory~\cite{nusspickel2023effective}, which might be even more critical in the case of (g)LPFET, where embedding clusters do not share the same chemical potential, in general. This should be investigated further in future work.


\subsection{Ab initio hydrogen chain}\label{Hydrogen_chain}

\begin{figure}
    \centering
    \includegraphics[width=1\linewidth]{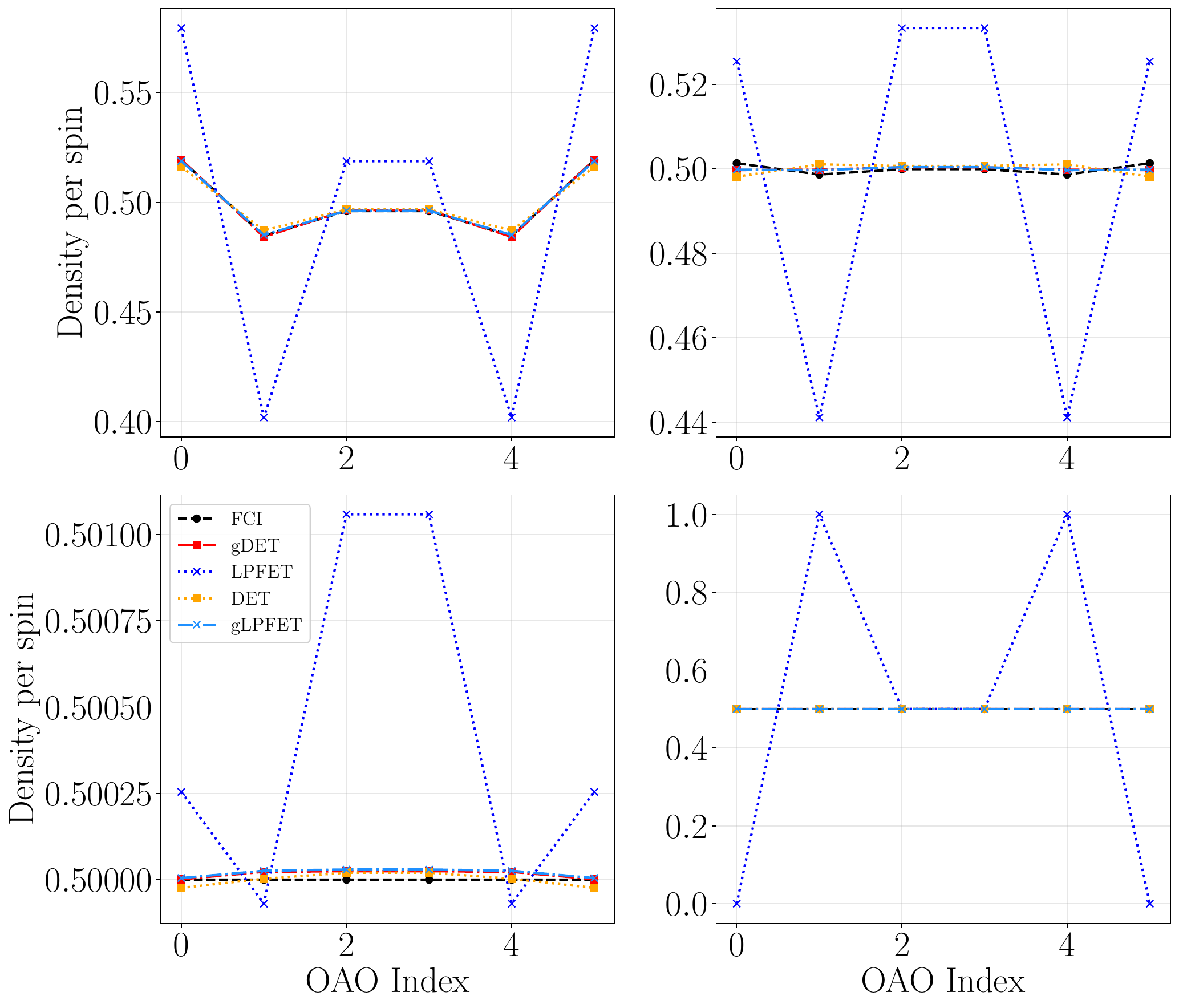}
    \caption{Density profiles (\ie, occupations of OAOs in this case) computed at the (g)LPFET and (g)DET levels of embedding for a chain of six equispaced hydrogen atoms with various bond distances. When not distinguishable, gLPFET is on top of gDET. Comparison is made with FCI. See text for further details.}
    \label{fig:H_Ring_density}
\end{figure}

The following discussion analyzes the performance of gLPFET  and gDET on an {\it ab initio} system which is a chain of six equispaced hydrogen atoms. The strength of electron correlation is modulated in this case by varying the bond distance $R$. While being weakly correlated at equilibrium ($R\approx 0.9$ \AA), stretching makes the molecule strongly correlated. Fig.~\ref{fig:H_Ring_density} shows the localized orbital occupancies obtained from (g)LPFET  and (g)DET, compared with the exact FCI ones, for bond lengths ranging from $R=0.9$ to $3.5$ \AA. The failure of LPFET at equilibrium, which was reported previously in Ref.~\cite{LPFET}, contrasts with the high accuracy of gLPFET. As expected for an {\it ab initio} system and unlike in the Hubbard ring, DET and gDET are not identical (in this case, the Hx potential of gDET is non-local in the localized orbital basis). They give similar (and relatively accurate) results though. For the larger $R=1.5$~\AA~bond distance, the FCI density profile is nearly uniform, a feature that both gLPFET and gDET (which are on top of each other) reproduce very well. While DET does too, but less accurately, the LPFET density profile is still dramatically wrong. When the chain approaches the dissociation limit, for $R=3.0$ \AA, LPFET becomes qualitatively correct, as expected~\cite{LPFET}, but it cannot compete with gLPFET and (g)DET, which are very accurate in this case. Note that, for the larger $R=3.5$ \AA~bond distance, LPFET's prediction of the electronic structure is completely wrong (electrons leave the edges of the chain) while gLPFET and (g)DET converge to the expected uniform density profile. Moreover, in contrast to LPFET, which fails to converge at some large bond distances~\cite{LPFET}, gLPFET remains numerically stable as $R$ increases.\\ 

\begin{figure}
    \centering
    \includegraphics[width=1.6\linewidth]{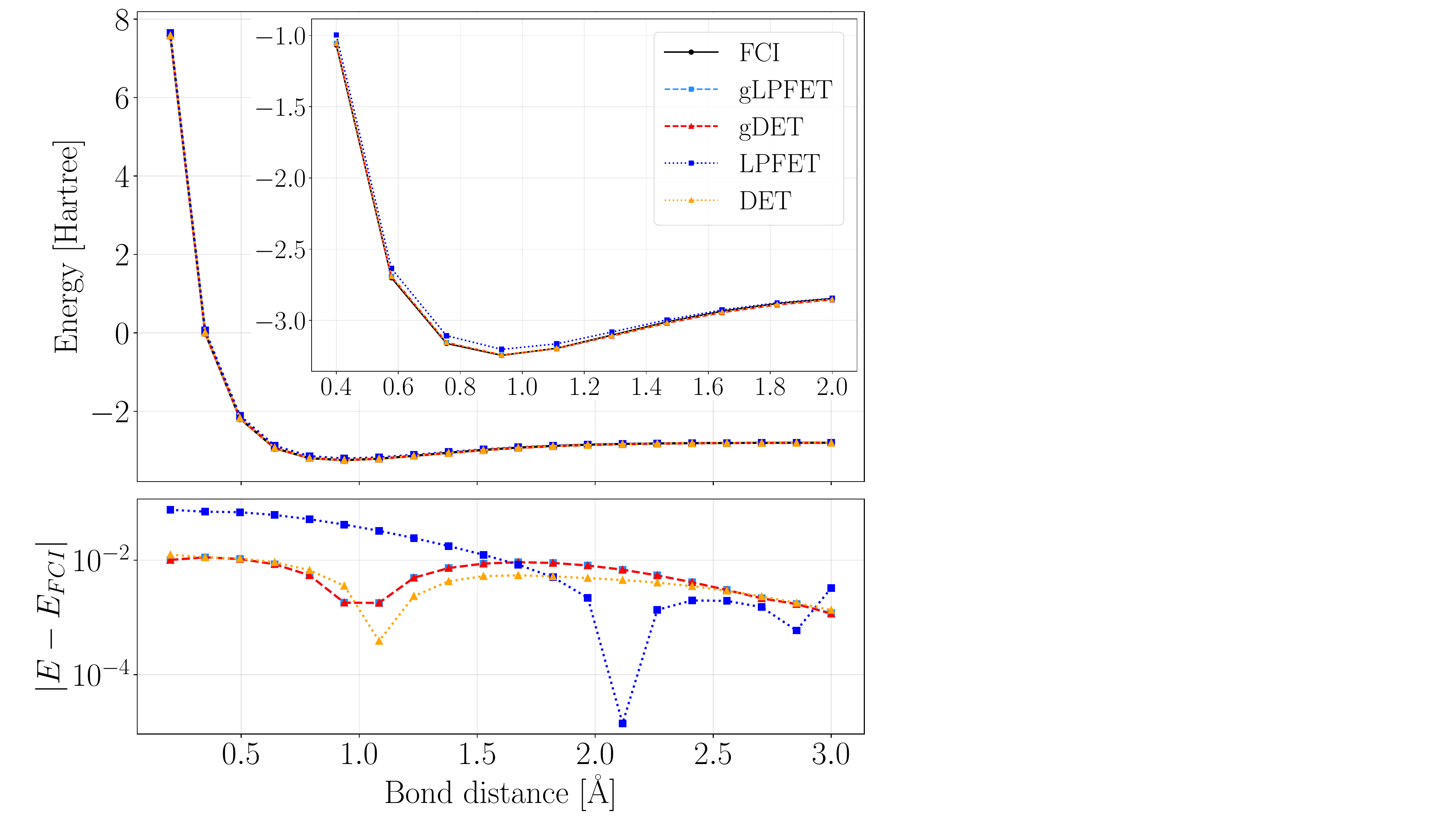}
    \caption{Potential energy curve of the hydrogen chain computed at the (g)LPFET and (g)DET levels of embedding. Comparison is made with FCI. See text for further details.}
    \label{fig:H_Ring_Energy}
\end{figure}

For completeness, we show in Fig.~\ref{fig:H_Ring_Energy} the potential energy curves obtained with the different embedding methods. Like for the Hubbard ring, a democratic evaluation of the correlated 1- and 2RDMs has been used for calculating the total energies. Unlike LPFET, whose dramatic failure in weakly correlated regimes translates here into a substantial underestimation of the correlation energy around the equilibrium geometry, gLPFET and (g)DET remain relatively close to the FCI benchmark at all bond distances. In this system, the accuracy of gLPFET in reproducing the FCI density profile is reflected also in the calculation of the energy.


\section{Conclusions and perspectives}\label{sec:conclusions}

A so-called generalized flavor (gLPFET) of the LPFET derived recently by the authors~\cite{LPFET} has been introduced, rationalized, and tested successfully on model systems. While the original LPFET relies on a full-size lattice KS system (the lattice consisting of localized orbitals for an {\it ab initio} Hamiltonian) where the Hxc potential is local, in gLPFET, the embedding clusters are constructed from a gKS system instead, where the Hx potential is that of a HF calculation, complemented by a local correlation potential that is determined self-consistently. As we use a (non-local) Hx potential that is not exclusively expressed in terms of the density (\ie, the localized orbital occupancies in this context) but in terms of the 1RDM instead, the reference system the embedding relies on is not a regular lattice KS one, hence the name {\it generalized}. This strategy allows for a better description of longer-range interactions, in principle. Most importantly, the expression of the LPFET impurity chemical potential, from which important features of strongly correlated regimes can be captured~\cite{LPFET}, has been adapted to gLPFET, with a proper rationale derived in Appendix~\ref{appendix:rationale_gLPFET}. The final key expression, given in Eq.~(\ref{eq:ind_LPFET_chem_pot}), is trivially obtained from that of LPFET by removing the local Hx potential contribution. As expected from our derivations and confirmed numerically on test systems, the gLPFET ansatz is very accurate in strongly correlated regimes. It captures the same physics (that conventional (g)DET misses) as LPFET while being numerically stable. Moreover, and this is the good surprize of this work, it completely fixes the flaw of LPFET in the weakly correlated regime, by {\it not} involving the Hx potential explicitly in the optimization of the impurity chemical potential.\\

Despite these very promising results, various formal and practical questions still need to be addressed. First of all, in the light of Ref.~\cite{LPFET}, we expect gLPFET to be the approximation to an in-principle exact (lattice) density-functional embedding theory (DFET) based on gKS-DFT. Such a derivation, which is left for future work, may serve as a reference for improving the method systematically in more complex systems. Recent findings in DFET~\cite{Wesolowski25_Density}, when it is formulated in the continuous real space, should also be taken into consideration in this task. Another critical point for molecular systems, in particular, is the calculation of total energies and the related $N$-representability problem~\cite{nusspickel2023effective}. Combining the gLPFET formalism with that of the recently developed Ghost Gutzwiller approach~\cite{mejuto2024quantum,lanata2023derivation,giuli2025linearfoundationmodelquantum} is appealing in this respect. Work in these different directions is currently in progress.

\section*{Acknowledgements}
This work of the Interdisciplinary Thematic Institute QMat, as part of the ITI 2021‐2028 program of the University of Strasbourg, CNRS and Inserm, was supported by IdEx Unistra (ANR 10 IDEX 0002), and by SFRI STRAT’US project (ANR 20 SFRI 0012) and EUR QMAT ANR‐17‐EURE‐0024 under the framework of the French Investments for the Future Program. This work benefited also from State support managed by the National Research Agency under the France 2030 program, referenced by ANR‐22‐CMAS‐0001 as well as ANR-23-PETQ-0006.

\appendix

\section{Impurity chemical potential expression adapted to a full-size gKS reference}\label{appendix:rationale_gLPFET}

By analogy with Appendix~C of Ref.~\citenum{LPFET}, we consider the following local density approximation to the exact correlation functional of gKS-DFT [its definition for $N$-representable densities is given in Eq.~(\ref{eq:corr_func_gKS-DFT})],
\be\label{eq:lda_gKS}
\begin{split}
E_{\rm c}({\bf n})\approx E^{(i)}_{\rm c}(n_i)&=\innerop{\Psi^{(i)}(n_i)}{\hat{T}+\hat{U}}{\Psi^{(i)}(n_i)}
\\
&\quad
-\innerop{\Phi^{(i)}(n_i)}{\hat{T}+\hat{U}}{\Phi^{(i)}(n_i)}
,
\end{split}
\ee
so that we can approximate the correlation potential as follows, according to Eqs.~(\ref{eq:exact_corr_pot_gKS}) and (\ref{eq:exact_dens_mapping_gKS}),
\be\label{eq:LDA_corr_pot_expression}
v_i^{\rm c}\approx \left.\dfrac{\partial E^{(i)}_{\rm c}(n_i)}{\partial n_i}\right|_{n_i=n_i^{\Psi_0}}. 
\ee
In the expression of Eq.~(\ref{eq:lda_gKS}), which is expected to be accurate when electron correlation is strong and local, $\Psi^{(i)}(n_i)$ is the ground state of the interacting local density functional embedding cluster's Hamiltonian
\be\label{eq:local_dens_func_int_cluster_Hamil}
\hat{\mathcal{H}}^{(i)}(n_i)=\hat{P}^{(i)}\hat{H}\hat{P}^{(i)}-\mu^{(i)}_{\rm imp}(n_i)\hat{n}_i,
\ee
where the impurity chemical potential is adjusted such that the local density $n_i$ is reproduced, \ie,
\be\label{eq:int_dens_constraint_cluster}
\langle \Psi^{(i)}(n_i)\vert \hat{n}_i\vert \Psi^{(i)}(n_i)\rangle=n_i.
\ee
Unlike in regular LPFET~\cite{LPFET}, which relies on a non-interacting KS system, the Slater determinant $\Phi^{(i)}(n_i)$ in Eq.~(\ref{eq:lda_gKS}) is defined as the HF approximation to a modified interacting embedding cluster's Hamiltonian,
\be\label{eq:HF_applied_to_int_cluster_with_add_chem_pot}
\begin{split}
&\Phi^{(i)}(n_i)
\\
&=\argmin_{\Phi^{(i)}}\innerop{\Phi^{(i)}}{\hat{P}^{(i)}\hat{H}^{\rm c}\hat{P}^{(i)}-\mu^{(i)}_{\rm
gKS}(n_i)\hat{n}_i}{\Phi^{(i)}},
\end{split}
\ee
where
\be
\hat{H}^{\rm c}:=\hat{H}+\hat{V}^{\rm c}
\ee
incorporates the local correlation potential operator $\hat{V}^{\rm c}=\sum_{k}v^{\rm c}_k\hat{n}_k$, and the gKS analog of the impurity chemical potential $\mu^{(i)}_{\rm
gKS}(n_i)$ is adjusted such that the exact same local density is reproduced,
\be\label{eq:gKS_dens_constraint_cluster}
\langle \Phi^{(i)}(n_i)\vert \hat{n}_i\vert \Phi^{(i)}(n_i)\rangle=n_i.
\ee
At this point it is important to realize that applying the variational principle of Eq.~(\ref{eq:VP_gKS-DFT}), which leads to the (self-consistent) diagonalization of the full-size gKS Hamiltonian (see Eq.~(\ref{eq:sc_gKS_equation})), is equivalent to performing the minimizing of Eq.~(\ref{eq:HF_applied_to_int_cluster_with_add_chem_pot}) without any additional gKS chemical potential. The reason is that the clusterization procedure is exact for any idempotent (one-electron reduced) density matrix~\cite{sekaran2023unified}, like that of the reference gKS Slater determinant, from which the projector $\hat{P}^{(i)}$ is determined. As both approaches should reproduce, in the exact theory, the true physical density profile $\underline{n}^{\Psi_0}$, we can conclude that
\be\label{eq:no_gKS_chem_pot_exact_dens}
\left.\mu^{(i)}_{\rm
gKS}(n_i)\right|_{n_i=n_i^{\Psi_0}}{=}0.
\ee
Let us now proceed with the evaluation of the density-functional derivative in Eq.~(\ref{eq:LDA_corr_pot_expression}). According to the density constraints of Eqs.~(\ref{eq:int_dens_constraint_cluster}) and (\ref{eq:gKS_dens_constraint_cluster}), and the fact that the densities on the (non-overlapping) impurity and bath orbitals sum up to 2 (in the present single-orbital embedding case~\cite{LPFET}), the local external potential operator can be added to both terms on the right-hand side of Eq.~(\ref{eq:lda_gKS}), so that the local correlation density functional can be rewritten as follows,  
\be
\begin{split}
E^{(i)}_{\rm c}(n_i)&=\innerop{\Psi^{(i)}(n_i)}{\hat{P}^{(i)}\hat{H}\hat{P}^{(i)}}{\Psi^{(i)}(n_i)}
\\
&\quad
-\innerop{\Phi^{(i)}(n_i)}{\hat{P}^{(i)}\hat{H}\hat{P}^{(i)}}{\Phi^{(i)}(n_i)},
\end{split}
\ee
or, equivalently, if we incorporate (and subtract) the local correlation potential operator, 
\be\label{eq:lda_corr_func_3_terms}
\begin{split}
E^{(i)}_{\rm c}(n_i)&=\innerop{\Psi^{(i)}(n_i)}{\hat{P}^{(i)}\hat{H}\hat{P}^{(i)}}{\Psi^{(i)}(n_i)}
\\
&\quad
-\innerop{\Phi^{(i)}(n_i)}{\hat{P}^{(i)}\hat{H}^{\rm c}\hat{P}^{(i)}}{\Phi^{(i)}(n_i)}
\\
&\quad
+\innerop{\Phi^{(i)}(n_i)}{\hat{P}^{(i)}\hat{V}^{\rm c}\hat{P}^{(i)}}{\Phi^{(i)}(n_i)}
.
\end{split}
\ee
When applied to Eq.~(\ref{eq:HF_applied_to_int_cluster_with_add_chem_pot}), the Hellmann--Feynman theorem reads
\be
\begin{split}
&\dfrac{\partial}{\partial n_i}\langle{\Phi^{(i)}(n_i)}\vert{\hat{P}^{(i)}\hat{H}^{\rm c}\hat{P}^{(i)}-\mu^{(i)}_{\rm
gKS}(n_i)\hat{n}_i}\vert{\Phi^{(i)}(n_i)}\rangle
\\
&=-\left({\partial \mu^{(i)}_{\rm
gKS}(n_i)}/{\partial n_i}\right)n_i,
\end{split}
\ee
so that, according to Eq.~(\ref{eq:no_gKS_chem_pot_exact_dens}), 
\be
\begin{split}
&\left.\dfrac{\partial}{\partial n_i}\langle{\Phi^{(i)}(n_i)}\vert\hat{P}^{(i)}\hat{H}^{\rm c}\hat{P}^{(i)}\vert{\Phi^{(i)}(n_i)}\rangle\right|_{n_i=n_i^{\Psi_0}}
\\
&=\mu^{(i)}_{\rm
gKS}(n_i^{\Psi_0})=0.
\end{split}
\ee
Thus, we deduce from Eqs.~(\ref{eq:local_dens_func_int_cluster_Hamil}) and (\ref{eq:lda_corr_func_3_terms}) the following expression for the local density-functional correlation potential, 
\be\label{eq:final_exp_lda_corr_pot}
\left.\dfrac{\partial E^{(i)}_{\rm c}(n_i)}{\partial n_i}\right|_{n_i=n_i^{\Psi_0}}=\mu^{(i)}_{\rm imp}
+v^{\rm c}_i-\langle b^{(i)}\vert \hat{v}^{\rm c}\vert b^{(i)}\rangle, 
\ee
where $\hat{v}^{\rm c}$ is the (local) single-electron  correlation potential operator and
\be
\begin{split}
\langle b^{(i)}\vert \hat{v}^{\rm c}\vert b^{(i)}\rangle&=
\sum_{kl}\langle b^{(i)}\vert \chi_k\rangle\langle \chi_k\vert \hat{v}^{\rm c}\vert \chi_l\rangle\langle \chi_l\vert b^{(i)}\rangle   
\\
&=\sum_k \inner{b^{(i)}}{\chi_k}^2v^{\rm c}_k.
\end{split}
\ee
The final expression of the impurity chemical potential given in Eq.~(\ref{eq:ind_LPFET_chem_pot}) is straightforwardly recovered from our initial local density approximation (see Eq.~(\ref{eq:LDA_corr_pot_expression})) by inserting the above expression into Eq.~(\ref{eq:final_exp_lda_corr_pot}).




\begin{thebibliography}{42}%
\makeatletter
\providecommand \@ifxundefined [1]{%
 \@ifx{#1\undefined}
}%
\providecommand \@ifnum [1]{%
 \ifnum #1\expandafter \@firstoftwo
 \else \expandafter \@secondoftwo
 \fi
}%
\providecommand \@ifx [1]{%
 \ifx #1\expandafter \@firstoftwo
 \else \expandafter \@secondoftwo
 \fi
}%
\providecommand \natexlab [1]{#1}%
\providecommand \enquote  [1]{``#1''}%
\providecommand \bibnamefont  [1]{#1}%
\providecommand \bibfnamefont [1]{#1}%
\providecommand \citenamefont [1]{#1}%
\providecommand \href@noop [0]{\@secondoftwo}%
\providecommand \href [0]{\begingroup \@sanitize@url \@href}%
\providecommand \@href[1]{\@@startlink{#1}\@@href}%
\providecommand \@@href[1]{\endgroup#1\@@endlink}%
\providecommand \@sanitize@url [0]{\catcode `\\12\catcode `\$12\catcode
  `\&12\catcode `\#12\catcode `\^12\catcode `\_12\catcode `\%12\relax}%
\providecommand \@@startlink[1]{}%
\providecommand \@@endlink[0]{}%
\providecommand \url  [0]{\begingroup\@sanitize@url \@url }%
\providecommand \@url [1]{\endgroup\@href {#1}{\urlprefix }}%
\providecommand \urlprefix  [0]{URL }%
\providecommand \Eprint [0]{\href }%
\providecommand \doibase [0]{http://dx.doi.org/}%
\providecommand \selectlanguage [0]{\@gobble}%
\providecommand \bibinfo  [0]{\@secondoftwo}%
\providecommand \bibfield  [0]{\@secondoftwo}%
\providecommand \translation [1]{[#1]}%
\providecommand \BibitemOpen [0]{}%
\providecommand \bibitemStop [0]{}%
\providecommand \bibitemNoStop [0]{.\EOS\space}%
\providecommand \EOS [0]{\spacefactor3000\relax}%
\providecommand \BibitemShut  [1]{\csname bibitem#1\endcsname}%
\let\auto@bib@innerbib\@empty
\bibitem [{\citenamefont {Bartlett}\ and\ \citenamefont
  {Musia\l{}}(2007)}]{RevModPhys.79.291}%
  \BibitemOpen
  \bibfield  {author} {\bibinfo {author} {\bibfnamefont {R.~J.}\ \bibnamefont
  {Bartlett}}\ and\ \bibinfo {author} {\bibfnamefont {M.}~\bibnamefont
  {Musia\l{}}},\ }\href {\doibase 10.1103/RevModPhys.79.291} {\bibfield
  {journal} {\bibinfo  {journal} {Rev. Mod. Phys.}\ }\textbf {\bibinfo {volume}
  {79}},\ \bibinfo {pages} {291} (\bibinfo {year} {2007})}\BibitemShut
  {NoStop}%
\bibitem [{\citenamefont {Alavi}\ \emph {et~al.}(2024)\citenamefont {Alavi},
  \citenamefont {Atalar}, \citenamefont {Berkelbach}, \citenamefont {Booth},
  \citenamefont {Chan}, \citenamefont {Evangelista}, \citenamefont {Goldzak},
  \citenamefont {Grüneis}, \citenamefont {Harsha}, \citenamefont {Kapil},
  \citenamefont {Knowles}, \citenamefont {Lepetit}, \citenamefont {Liebert},
  \citenamefont {Nejad}, \citenamefont {Neufeld}, \citenamefont {Novoa},
  \citenamefont {Pernal}, \citenamefont {Plasser}, \citenamefont {Rehman},
  \citenamefont {Shi}, \citenamefont {Tew}, \citenamefont {Wang}, \citenamefont
  {Mejuto-Zaera}, \citenamefont {Zgid}, \citenamefont {Zhu}, \citenamefont
  {Zhu},\ and\ \citenamefont {Zwijnenburg}}]{Alavi24_Faraday_Correlation}%
  \BibitemOpen
  \bibfield  {author} {\bibinfo {author} {\bibfnamefont {A.}~\bibnamefont
  {Alavi}}, \bibinfo {author} {\bibfnamefont {K.}~\bibnamefont {Atalar}},
  \bibinfo {author} {\bibfnamefont {T.~C.}\ \bibnamefont {Berkelbach}},
  \bibinfo {author} {\bibfnamefont {G.~H.}\ \bibnamefont {Booth}}, \bibinfo
  {author} {\bibfnamefont {G.~K.-L.}\ \bibnamefont {Chan}}, \bibinfo {author}
  {\bibfnamefont {F.~A.}\ \bibnamefont {Evangelista}}, \bibinfo {author}
  {\bibfnamefont {T.}~\bibnamefont {Goldzak}}, \bibinfo {author} {\bibfnamefont
  {A.}~\bibnamefont {Grüneis}}, \bibinfo {author} {\bibfnamefont
  {G.}~\bibnamefont {Harsha}}, \bibinfo {author} {\bibfnamefont
  {V.}~\bibnamefont {Kapil}}, \bibinfo {author} {\bibfnamefont
  {P.}~\bibnamefont {Knowles}}, \bibinfo {author} {\bibfnamefont {M.-B.}\
  \bibnamefont {Lepetit}}, \bibinfo {author} {\bibfnamefont {J.}~\bibnamefont
  {Liebert}}, \bibinfo {author} {\bibfnamefont {A.}~\bibnamefont {Nejad}},
  \bibinfo {author} {\bibfnamefont {V.~A.}\ \bibnamefont {Neufeld}}, \bibinfo
  {author} {\bibfnamefont {T.}~\bibnamefont {Novoa}}, \bibinfo {author}
  {\bibfnamefont {K.}~\bibnamefont {Pernal}}, \bibinfo {author} {\bibfnamefont
  {F.}~\bibnamefont {Plasser}}, \bibinfo {author} {\bibfnamefont
  {U.}~\bibnamefont {Rehman}}, \bibinfo {author} {\bibfnamefont {B.~X.}\
  \bibnamefont {Shi}}, \bibinfo {author} {\bibfnamefont {D.~P.}\ \bibnamefont
  {Tew}}, \bibinfo {author} {\bibfnamefont {Z.}~\bibnamefont {Wang}}, \bibinfo
  {author} {\bibfnamefont {C.}~\bibnamefont {Mejuto-Zaera}}, \bibinfo {author}
  {\bibfnamefont {D.}~\bibnamefont {Zgid}}, \bibinfo {author} {\bibfnamefont
  {A.}~\bibnamefont {Zhu}}, \bibinfo {author} {\bibfnamefont {T.}~\bibnamefont
  {Zhu}}, \ and\ \bibinfo {author} {\bibfnamefont {M.~A.}\ \bibnamefont
  {Zwijnenburg}},\ }\href {\doibase 10.1039/D4FD90044H} {\bibfield  {journal}
  {\bibinfo  {journal} {Faraday Discuss.}\ }\textbf {\bibinfo {volume} {254}},\
  \bibinfo {pages} {682} (\bibinfo {year} {2024})}\BibitemShut {NoStop}%
\bibitem [{\citenamefont {Teale}\ \emph {et~al.}(2022)\citenamefont {Teale},
  \citenamefont {Helgaker}, \citenamefont {Savin}, \citenamefont {Adamo},
  \citenamefont {Aradi}, \citenamefont {Arbuznikov}, \citenamefont {Ayers},
  \citenamefont {Baerends}, \citenamefont {Barone}, \citenamefont {Calaminici},
  \citenamefont {Canc{\`e}s}, \citenamefont {Carter}, \citenamefont
  {Chattaraj}, \citenamefont {Chermette}, \citenamefont {Ciofini},
  \citenamefont {Crawford}, \citenamefont {De~Proft}, \citenamefont {Dobson},
  \citenamefont {Draxl}, \citenamefont {Frauenheim}, \citenamefont {Fromager},
  \citenamefont {Fuentealba}, \citenamefont {Gagliardi}, \citenamefont {Galli},
  \citenamefont {Gao}, \citenamefont {Geerlings}, \citenamefont {Gidopoulos},
  \citenamefont {Gill}, \citenamefont {Gori-Giorgi}, \citenamefont
  {G{\"o}rling}, \citenamefont {Gould}, \citenamefont {Grimme}, \citenamefont
  {Gritsenko}, \citenamefont {Jensen}, \citenamefont {Johnson}, \citenamefont
  {Jones}, \citenamefont {Kaupp}, \citenamefont {K{\"o}ster}, \citenamefont
  {Kronik}, \citenamefont {Krylov}, \citenamefont {Kvaal}, \citenamefont
  {Laestadius}, \citenamefont {Levy}, \citenamefont {Lewin}, \citenamefont
  {Liu}, \citenamefont {Loos}, \citenamefont {Maitra}, \citenamefont {Neese},
  \citenamefont {Perdew}, \citenamefont {Pernal}, \citenamefont {Pernot},
  \citenamefont {Piecuch}, \citenamefont {Rebolini}, \citenamefont {Reining},
  \citenamefont {Romaniello}, \citenamefont {Ruzsinszky}, \citenamefont
  {Salahub}, \citenamefont {Scheffler}, \citenamefont {Schwerdtfeger},
  \citenamefont {Staroverov}, \citenamefont {Sun}, \citenamefont {Tellgren},
  \citenamefont {Tozer}, \citenamefont {Trickey}, \citenamefont {Ullrich},
  \citenamefont {Vela}, \citenamefont {Vignale}, \citenamefont {Wesolowski},
  \citenamefont {Xu},\ and\ \citenamefont {Yang}}]{Teale2022_DFT_exchange}%
  \BibitemOpen
  \bibfield  {author} {\bibinfo {author} {\bibfnamefont {A.~M.}\ \bibnamefont
  {Teale}}, \bibinfo {author} {\bibfnamefont {T.}~\bibnamefont {Helgaker}},
  \bibinfo {author} {\bibfnamefont {A.}~\bibnamefont {Savin}}, \bibinfo
  {author} {\bibfnamefont {C.}~\bibnamefont {Adamo}}, \bibinfo {author}
  {\bibfnamefont {B.}~\bibnamefont {Aradi}}, \bibinfo {author} {\bibfnamefont
  {A.~V.}\ \bibnamefont {Arbuznikov}}, \bibinfo {author} {\bibfnamefont
  {P.~W.}\ \bibnamefont {Ayers}}, \bibinfo {author} {\bibfnamefont {E.~J.}\
  \bibnamefont {Baerends}}, \bibinfo {author} {\bibfnamefont {V.}~\bibnamefont
  {Barone}}, \bibinfo {author} {\bibfnamefont {P.}~\bibnamefont {Calaminici}},
  \bibinfo {author} {\bibfnamefont {E.}~\bibnamefont {Canc{\`e}s}}, \bibinfo
  {author} {\bibfnamefont {E.~A.}\ \bibnamefont {Carter}}, \bibinfo {author}
  {\bibfnamefont {P.~K.}\ \bibnamefont {Chattaraj}}, \bibinfo {author}
  {\bibfnamefont {H.}~\bibnamefont {Chermette}}, \bibinfo {author}
  {\bibfnamefont {I.}~\bibnamefont {Ciofini}}, \bibinfo {author} {\bibfnamefont
  {T.~D.}\ \bibnamefont {Crawford}}, \bibinfo {author} {\bibfnamefont
  {F.}~\bibnamefont {De~Proft}}, \bibinfo {author} {\bibfnamefont {J.~F.}\
  \bibnamefont {Dobson}}, \bibinfo {author} {\bibfnamefont {C.}~\bibnamefont
  {Draxl}}, \bibinfo {author} {\bibfnamefont {T.}~\bibnamefont {Frauenheim}},
  \bibinfo {author} {\bibfnamefont {E.}~\bibnamefont {Fromager}}, \bibinfo
  {author} {\bibfnamefont {P.}~\bibnamefont {Fuentealba}}, \bibinfo {author}
  {\bibfnamefont {L.}~\bibnamefont {Gagliardi}}, \bibinfo {author}
  {\bibfnamefont {G.}~\bibnamefont {Galli}}, \bibinfo {author} {\bibfnamefont
  {J.}~\bibnamefont {Gao}}, \bibinfo {author} {\bibfnamefont {P.}~\bibnamefont
  {Geerlings}}, \bibinfo {author} {\bibfnamefont {N.}~\bibnamefont
  {Gidopoulos}}, \bibinfo {author} {\bibfnamefont {P.~M.~W.}\ \bibnamefont
  {Gill}}, \bibinfo {author} {\bibfnamefont {P.}~\bibnamefont {Gori-Giorgi}},
  \bibinfo {author} {\bibfnamefont {A.}~\bibnamefont {G{\"o}rling}}, \bibinfo
  {author} {\bibfnamefont {T.}~\bibnamefont {Gould}}, \bibinfo {author}
  {\bibfnamefont {S.}~\bibnamefont {Grimme}}, \bibinfo {author} {\bibfnamefont
  {O.}~\bibnamefont {Gritsenko}}, \bibinfo {author} {\bibfnamefont {H.~J.~A.}\
  \bibnamefont {Jensen}}, \bibinfo {author} {\bibfnamefont {E.~R.}\
  \bibnamefont {Johnson}}, \bibinfo {author} {\bibfnamefont {R.~O.}\
  \bibnamefont {Jones}}, \bibinfo {author} {\bibfnamefont {M.}~\bibnamefont
  {Kaupp}}, \bibinfo {author} {\bibfnamefont {A.~M.}\ \bibnamefont
  {K{\"o}ster}}, \bibinfo {author} {\bibfnamefont {L.}~\bibnamefont {Kronik}},
  \bibinfo {author} {\bibfnamefont {A.~I.}\ \bibnamefont {Krylov}}, \bibinfo
  {author} {\bibfnamefont {S.}~\bibnamefont {Kvaal}}, \bibinfo {author}
  {\bibfnamefont {A.}~\bibnamefont {Laestadius}}, \bibinfo {author}
  {\bibfnamefont {M.}~\bibnamefont {Levy}}, \bibinfo {author} {\bibfnamefont
  {M.}~\bibnamefont {Lewin}}, \bibinfo {author} {\bibfnamefont
  {S.}~\bibnamefont {Liu}}, \bibinfo {author} {\bibfnamefont {P.-F.}\
  \bibnamefont {Loos}}, \bibinfo {author} {\bibfnamefont {N.~T.}\ \bibnamefont
  {Maitra}}, \bibinfo {author} {\bibfnamefont {F.}~\bibnamefont {Neese}},
  \bibinfo {author} {\bibfnamefont {J.~P.}\ \bibnamefont {Perdew}}, \bibinfo
  {author} {\bibfnamefont {K.}~\bibnamefont {Pernal}}, \bibinfo {author}
  {\bibfnamefont {P.}~\bibnamefont {Pernot}}, \bibinfo {author} {\bibfnamefont
  {P.}~\bibnamefont {Piecuch}}, \bibinfo {author} {\bibfnamefont
  {E.}~\bibnamefont {Rebolini}}, \bibinfo {author} {\bibfnamefont
  {L.}~\bibnamefont {Reining}}, \bibinfo {author} {\bibfnamefont
  {P.}~\bibnamefont {Romaniello}}, \bibinfo {author} {\bibfnamefont
  {A.}~\bibnamefont {Ruzsinszky}}, \bibinfo {author} {\bibfnamefont {D.~R.}\
  \bibnamefont {Salahub}}, \bibinfo {author} {\bibfnamefont {M.}~\bibnamefont
  {Scheffler}}, \bibinfo {author} {\bibfnamefont {P.}~\bibnamefont
  {Schwerdtfeger}}, \bibinfo {author} {\bibfnamefont {V.~N.}\ \bibnamefont
  {Staroverov}}, \bibinfo {author} {\bibfnamefont {J.}~\bibnamefont {Sun}},
  \bibinfo {author} {\bibfnamefont {E.}~\bibnamefont {Tellgren}}, \bibinfo
  {author} {\bibfnamefont {D.~J.}\ \bibnamefont {Tozer}}, \bibinfo {author}
  {\bibfnamefont {S.~B.}\ \bibnamefont {Trickey}}, \bibinfo {author}
  {\bibfnamefont {C.~A.}\ \bibnamefont {Ullrich}}, \bibinfo {author}
  {\bibfnamefont {A.}~\bibnamefont {Vela}}, \bibinfo {author} {\bibfnamefont
  {G.}~\bibnamefont {Vignale}}, \bibinfo {author} {\bibfnamefont {T.~A.}\
  \bibnamefont {Wesolowski}}, \bibinfo {author} {\bibfnamefont
  {X.}~\bibnamefont {Xu}}, \ and\ \bibinfo {author} {\bibfnamefont
  {W.}~\bibnamefont {Yang}},\ }\href {\doibase 10.1039/D2CP02827A} {\bibfield
  {journal} {\bibinfo  {journal} {Phys. Chem. Chem. Phys.}\ }\textbf {\bibinfo
  {volume} {24}},\ \bibinfo {pages} {28700} (\bibinfo {year}
  {2022})}\BibitemShut {NoStop}%
\bibitem [{\citenamefont {Sun}\ and\ \citenamefont
  {Chan}(2016)}]{sun2016quantum}%
  \BibitemOpen
  \bibfield  {author} {\bibinfo {author} {\bibfnamefont {Q.}~\bibnamefont
  {Sun}}\ and\ \bibinfo {author} {\bibfnamefont {G.~K.-L.}\ \bibnamefont
  {Chan}},\ }\href {https://doi.org/10.1021/acs.accounts.6b00356} {\bibfield
  {journal} {\bibinfo  {journal} {Acc. Chem. Res.}\ }\textbf {\bibinfo {volume}
  {49}},\ \bibinfo {pages} {2705} (\bibinfo {year} {2016})}\BibitemShut
  {NoStop}%
\bibitem [{\citenamefont {Wasserman}\ and\ \citenamefont
  {Pavanello}(2020)}]{wasserman2020quantum}%
  \BibitemOpen
  \bibfield  {author} {\bibinfo {author} {\bibfnamefont {A.}~\bibnamefont
  {Wasserman}}\ and\ \bibinfo {author} {\bibfnamefont {M.}~\bibnamefont
  {Pavanello}},\ }\href {https://doi.org/10.1002/qua.26495} {\ \textbf
  {\bibinfo {volume} {120}},\ \bibinfo {pages} {e26495} (\bibinfo {year}
  {2020})}\BibitemShut {NoStop}%
\bibitem [{\citenamefont {Mejuto-Zaera}(2024)}]{mejuto2024quantum}%
  \BibitemOpen
  \bibfield  {author} {\bibinfo {author} {\bibfnamefont {C.}~\bibnamefont
  {Mejuto-Zaera}},\ }\href {http://dx.doi.org/10.1039/D4FD00053F} {\bibfield
  {journal} {\bibinfo  {journal} {Faraday Discuss.}\ }\textbf {\bibinfo
  {volume} {254}},\ \bibinfo {pages} {653} (\bibinfo {year}
  {2024})}\BibitemShut {NoStop}%
\bibitem [{\citenamefont {Evangelista}(2024)}]{Evangelista_2025_Concluding}%
  \BibitemOpen
  \bibfield  {author} {\bibinfo {author} {\bibfnamefont {F.~A.}\ \bibnamefont
  {Evangelista}},\ }\href {\doibase 10.1039/D4FD00152D} {\bibfield  {journal}
  {\bibinfo  {journal} {Faraday Discuss.}\ }\textbf {\bibinfo {volume} {254}},\
  \bibinfo {pages} {708} (\bibinfo {year} {2024})}\BibitemShut {NoStop}%
\bibitem [{\citenamefont {Verma}\ \emph {et~al.}(2026)\citenamefont {Verma},
  \citenamefont {Mitra}, \citenamefont {Wang}, \citenamefont {D’Cunha},
  \citenamefont {Jangid}, \citenamefont {Hennefarth}, \citenamefont {Agarawal},
  \citenamefont {Otis}, \citenamefont {Haldar}, \citenamefont {Hermes},\ and\
  \citenamefont {Gagliardi}}]{Verma26_Multireference_Embedding}%
  \BibitemOpen
  \bibfield  {author} {\bibinfo {author} {\bibfnamefont {S.}~\bibnamefont
  {Verma}}, \bibinfo {author} {\bibfnamefont {A.}~\bibnamefont {Mitra}},
  \bibinfo {author} {\bibfnamefont {Q.}~\bibnamefont {Wang}}, \bibinfo {author}
  {\bibfnamefont {R.}~\bibnamefont {D’Cunha}}, \bibinfo {author}
  {\bibfnamefont {B.}~\bibnamefont {Jangid}}, \bibinfo {author} {\bibfnamefont
  {M.~R.}\ \bibnamefont {Hennefarth}}, \bibinfo {author} {\bibfnamefont
  {V.}~\bibnamefont {Agarawal}}, \bibinfo {author} {\bibfnamefont
  {L.}~\bibnamefont {Otis}}, \bibinfo {author} {\bibfnamefont {S.}~\bibnamefont
  {Haldar}}, \bibinfo {author} {\bibfnamefont {M.~R.}\ \bibnamefont {Hermes}},
  \ and\ \bibinfo {author} {\bibfnamefont {L.}~\bibnamefont {Gagliardi}},\
  }\href {\doibase 10.1021/acs.chemrev.5c00486} {\bibfield  {journal} {\bibinfo
   {journal} {Chemical Reviews}\ }\textbf {\bibinfo {volume} {126}},\ \bibinfo
  {pages} {184} (\bibinfo {year} {2026})}\BibitemShut {NoStop}%
\bibitem [{\citenamefont {Lacombe}\ and\ \citenamefont
  {Maitra}(2020)}]{lacombe2020exactfac}%
  \BibitemOpen
  \bibfield  {author} {\bibinfo {author} {\bibfnamefont {L.}~\bibnamefont
  {Lacombe}}\ and\ \bibinfo {author} {\bibfnamefont {N.~T.}\ \bibnamefont
  {Maitra}},\ }\href {\doibase 10.1103/PhysRevLett.124.206401} {\bibfield
  {journal} {\bibinfo  {journal} {Phys. Rev. Lett.}\ }\textbf {\bibinfo
  {volume} {124}},\ \bibinfo {pages} {206401} (\bibinfo {year}
  {2020})}\BibitemShut {NoStop}%
\bibitem [{\citenamefont {Requist}\ and\ \citenamefont
  {Gross}(2021)}]{requist2021fset}%
  \BibitemOpen
  \bibfield  {author} {\bibinfo {author} {\bibfnamefont {R.}~\bibnamefont
  {Requist}}\ and\ \bibinfo {author} {\bibfnamefont {E.~K.~U.}\ \bibnamefont
  {Gross}},\ }\href {\doibase 10.1103/PhysRevLett.127.116401} {\bibfield
  {journal} {\bibinfo  {journal} {Phys. Rev. Lett.}\ }\textbf {\bibinfo
  {volume} {127}},\ \bibinfo {pages} {116401} (\bibinfo {year}
  {2021})}\BibitemShut {NoStop}%
\bibitem [{\citenamefont {Li}\ and\ \citenamefont
  {Zhu}(2024)}]{Jiachen24_Interacting-Bath}%
  \BibitemOpen
  \bibfield  {author} {\bibinfo {author} {\bibfnamefont {J.}~\bibnamefont
  {Li}}\ and\ \bibinfo {author} {\bibfnamefont {T.}~\bibnamefont {Zhu}},\
  }\href {\doibase 10.1103/PhysRevLett.133.216402} {\bibfield  {journal}
  {\bibinfo  {journal} {Phys. Rev. Lett.}\ }\textbf {\bibinfo {volume} {133}},\
  \bibinfo {pages} {216402} (\bibinfo {year} {2024})}\BibitemShut {NoStop}%
\bibitem [{\citenamefont {Knizia}\ and\ \citenamefont
  {Chan}(2012)}]{knizia2012density}%
  \BibitemOpen
  \bibfield  {author} {\bibinfo {author} {\bibfnamefont {G.}~\bibnamefont
  {Knizia}}\ and\ \bibinfo {author} {\bibfnamefont {G.~K.-L.}\ \bibnamefont
  {Chan}},\ }\href {\doibase 10.1103/PhysRevLett.109.186404} {\bibfield
  {journal} {\bibinfo  {journal} {Phys. Rev. Lett.}\ }\textbf {\bibinfo
  {volume} {109}},\ \bibinfo {pages} {186404} (\bibinfo {year}
  {2012})}\BibitemShut {NoStop}%
\bibitem [{\citenamefont {Knizia}\ and\ \citenamefont
  {Chan}(2013)}]{knizia2013density}%
  \BibitemOpen
  \bibfield  {author} {\bibinfo {author} {\bibfnamefont {G.}~\bibnamefont
  {Knizia}}\ and\ \bibinfo {author} {\bibfnamefont {G.~K.-L.}\ \bibnamefont
  {Chan}},\ }\href {\doibase 10.1021/ct301044e} {\bibfield  {journal} {\bibinfo
   {journal} {J. Chem. Theory Comput.}\ }\textbf {\bibinfo {volume} {9}},\
  \bibinfo {pages} {1428} (\bibinfo {year} {2013})}\BibitemShut {NoStop}%
\bibitem [{\citenamefont {Canc{\`e}s}\ \emph {et~al.}(2025)\citenamefont
  {Canc{\`e}s}, \citenamefont {Faulstich}, \citenamefont {Kirsch},
  \citenamefont {Letournel},\ and\ \citenamefont
  {Levitt}}]{cances2025analysis}%
  \BibitemOpen
  \bibfield  {author} {\bibinfo {author} {\bibfnamefont {E.}~\bibnamefont
  {Canc{\`e}s}}, \bibinfo {author} {\bibfnamefont {F.~M.}\ \bibnamefont
  {Faulstich}}, \bibinfo {author} {\bibfnamefont {A.}~\bibnamefont {Kirsch}},
  \bibinfo {author} {\bibfnamefont {E.}~\bibnamefont {Letournel}}, \ and\
  \bibinfo {author} {\bibfnamefont {A.}~\bibnamefont {Levitt}},\ }\href
  {https://doi.org/10.1002/cpa.22244} {\bibfield  {journal} {\bibinfo
  {journal} {Commun. Pure Appl. Math.}\ }\textbf {\bibinfo {volume} {78}},\
  \bibinfo {pages} {1359} (\bibinfo {year} {2025})}\BibitemShut {NoStop}%
\bibitem [{\citenamefont {Wouters}\ \emph {et~al.}(2017)\citenamefont
  {Wouters}, \citenamefont {A.~Jim{\'e}nez-Hoyos},\ and\ \citenamefont
  {KL~Chan}}]{wouters2017five}%
  \BibitemOpen
  \bibfield  {author} {\bibinfo {author} {\bibfnamefont {S.}~\bibnamefont
  {Wouters}}, \bibinfo {author} {\bibfnamefont {C.}~\bibnamefont
  {A.~Jim{\'e}nez-Hoyos}}, \ and\ \bibinfo {author} {\bibfnamefont
  {G.}~\bibnamefont {KL~Chan}},\ }\href
  {https://doi.org/10.1002/9781119129271.ch8} {\bibfield  {journal} {\bibinfo
  {journal} {Fragmentation: toward accurate calculations on complex molecular
  systems}\ ,\ \bibinfo {pages} {227}} (\bibinfo {year} {2017})}\BibitemShut
  {NoStop}%
\bibitem [{\citenamefont {Wouters}\ \emph {et~al.}(2016)\citenamefont
  {Wouters}, \citenamefont {Jim{\'e}nez-Hoyos}, \citenamefont {Sun},\ and\
  \citenamefont {Chan}}]{wouters2016practical}%
  \BibitemOpen
  \bibfield  {author} {\bibinfo {author} {\bibfnamefont {S.}~\bibnamefont
  {Wouters}}, \bibinfo {author} {\bibfnamefont {C.~A.}\ \bibnamefont
  {Jim{\'e}nez-Hoyos}}, \bibinfo {author} {\bibfnamefont {Q.}~\bibnamefont
  {Sun}}, \ and\ \bibinfo {author} {\bibfnamefont {G.~K.-L.}\ \bibnamefont
  {Chan}},\ }\href {\doibase 10.1021/acs.jctc.6b00316} {\bibfield  {journal}
  {\bibinfo  {journal} {J. Chem. Theory Comput.}\ }\textbf {\bibinfo {volume}
  {12}},\ \bibinfo {pages} {2706} (\bibinfo {year} {2016})}\BibitemShut
  {NoStop}%
\bibitem [{\citenamefont {Chen}\ \emph {et~al.}(2014)\citenamefont {Chen},
  \citenamefont {Booth}, \citenamefont {Sharma}, \citenamefont {Knizia},\ and\
  \citenamefont {Chan}}]{chen2014intermediate}%
  \BibitemOpen
  \bibfield  {author} {\bibinfo {author} {\bibfnamefont {Q.}~\bibnamefont
  {Chen}}, \bibinfo {author} {\bibfnamefont {G.~H.}\ \bibnamefont {Booth}},
  \bibinfo {author} {\bibfnamefont {S.}~\bibnamefont {Sharma}}, \bibinfo
  {author} {\bibfnamefont {G.}~\bibnamefont {Knizia}}, \ and\ \bibinfo {author}
  {\bibfnamefont {G.~K.-L.}\ \bibnamefont {Chan}},\ }\href {\doibase
  10.1103/PhysRevB.89.165134} {\bibfield  {journal} {\bibinfo  {journal} {Phys.
  Rev. B}\ }\textbf {\bibinfo {volume} {89}},\ \bibinfo {pages} {165134}
  (\bibinfo {year} {2014})}\BibitemShut {NoStop}%
\bibitem [{\citenamefont {Cui}\ \emph {et~al.}(2020)\citenamefont {Cui},
  \citenamefont {Sun}, \citenamefont {Ray}, \citenamefont {Zheng},
  \citenamefont {Sun},\ and\ \citenamefont {Chan}}]{cui2020ground}%
  \BibitemOpen
  \bibfield  {author} {\bibinfo {author} {\bibfnamefont {Z.-H.}\ \bibnamefont
  {Cui}}, \bibinfo {author} {\bibfnamefont {C.}~\bibnamefont {Sun}}, \bibinfo
  {author} {\bibfnamefont {U.}~\bibnamefont {Ray}}, \bibinfo {author}
  {\bibfnamefont {B.-X.}\ \bibnamefont {Zheng}}, \bibinfo {author}
  {\bibfnamefont {Q.}~\bibnamefont {Sun}}, \ and\ \bibinfo {author}
  {\bibfnamefont {G.~K.-L.}\ \bibnamefont {Chan}},\ }\href
  {https://doi.org/10.1103/PhysRevResearch.2.043259} {\bibfield  {journal}
  {\bibinfo  {journal} {Phys. Rev. Res.}\ }\textbf {\bibinfo {volume} {2}},\
  \bibinfo {pages} {043259} (\bibinfo {year} {2020})}\BibitemShut {NoStop}%
\bibitem [{\citenamefont {Kawano}\ and\ \citenamefont
  {Hotta}(2020)}]{kawano2020comparative}%
  \BibitemOpen
  \bibfield  {author} {\bibinfo {author} {\bibfnamefont {M.}~\bibnamefont
  {Kawano}}\ and\ \bibinfo {author} {\bibfnamefont {C.}~\bibnamefont {Hotta}},\
  }\href {https://doi.org/10.1103/PhysRevB.102.235111} {\bibfield  {journal}
  {\bibinfo  {journal} {Phys. Rev. B}\ }\textbf {\bibinfo {volume} {102}},\
  \bibinfo {pages} {235111} (\bibinfo {year} {2020})}\BibitemShut {NoStop}%
\bibitem [{\citenamefont {Mukherjee}\ and\ \citenamefont
  {Reichman}(2017)}]{mukherjee2017simple}%
  \BibitemOpen
  \bibfield  {author} {\bibinfo {author} {\bibfnamefont {S.}~\bibnamefont
  {Mukherjee}}\ and\ \bibinfo {author} {\bibfnamefont {D.~R.}\ \bibnamefont
  {Reichman}},\ }\href {https://doi.org/10.1103/PhysRevB.95.155111} {\bibfield
  {journal} {\bibinfo  {journal} {Phys. Rev. B}\ }\textbf {\bibinfo {volume}
  {95}},\ \bibinfo {pages} {155111} (\bibinfo {year} {2017})}\BibitemShut
  {NoStop}%
\bibitem [{\citenamefont {Sandhoefer}\ and\ \citenamefont
  {Chan}(2016)}]{sandhoefer2016density}%
  \BibitemOpen
  \bibfield  {author} {\bibinfo {author} {\bibfnamefont {B.}~\bibnamefont
  {Sandhoefer}}\ and\ \bibinfo {author} {\bibfnamefont {G.~K.-L.}\ \bibnamefont
  {Chan}},\ }\href {https://doi.org/10.1103/PhysRevB.94.085115} {\bibfield
  {journal} {\bibinfo  {journal} {Phys. Rev. B}\ }\textbf {\bibinfo {volume}
  {94}},\ \bibinfo {pages} {085115} (\bibinfo {year} {2016})}\BibitemShut
  {NoStop}%
\bibitem [{\citenamefont {Zheng}\ and\ \citenamefont
  {Chan}(2016)}]{zheng2016ground}%
  \BibitemOpen
  \bibfield  {author} {\bibinfo {author} {\bibfnamefont {B.-X.}\ \bibnamefont
  {Zheng}}\ and\ \bibinfo {author} {\bibfnamefont {G.~K.-L.}\ \bibnamefont
  {Chan}},\ }\href {https://doi.org/10.1103/PhysRevB.93.035126} {\bibfield
  {journal} {\bibinfo  {journal} {Phys. Rev. B}\ }\textbf {\bibinfo {volume}
  {93}},\ \bibinfo {pages} {035126} (\bibinfo {year} {2016})}\BibitemShut
  {NoStop}%
\bibitem [{\citenamefont {Zheng}\ \emph {et~al.}(2017)\citenamefont {Zheng},
  \citenamefont {Chung}, \citenamefont {Corboz}, \citenamefont {Ehlers},
  \citenamefont {Qin}, \citenamefont {Noack}, \citenamefont {Shi},
  \citenamefont {White}, \citenamefont {Zhang},\ and\ \citenamefont
  {Chan}}]{zheng2017stripe}%
  \BibitemOpen
  \bibfield  {author} {\bibinfo {author} {\bibfnamefont {B.-X.}\ \bibnamefont
  {Zheng}}, \bibinfo {author} {\bibfnamefont {C.-M.}\ \bibnamefont {Chung}},
  \bibinfo {author} {\bibfnamefont {P.}~\bibnamefont {Corboz}}, \bibinfo
  {author} {\bibfnamefont {G.}~\bibnamefont {Ehlers}}, \bibinfo {author}
  {\bibfnamefont {M.-P.}\ \bibnamefont {Qin}}, \bibinfo {author} {\bibfnamefont
  {R.~M.}\ \bibnamefont {Noack}}, \bibinfo {author} {\bibfnamefont
  {H.}~\bibnamefont {Shi}}, \bibinfo {author} {\bibfnamefont {S.~R.}\
  \bibnamefont {White}}, \bibinfo {author} {\bibfnamefont {S.}~\bibnamefont
  {Zhang}}, \ and\ \bibinfo {author} {\bibfnamefont {G.~K.-L.}\ \bibnamefont
  {Chan}},\ }\href {\doibase 10.1126/science.aam7127} {\bibfield  {journal}
  {\bibinfo  {journal} {Science}\ }\textbf {\bibinfo {volume} {358}},\ \bibinfo
  {pages} {1155} (\bibinfo {year} {2017})}\BibitemShut {NoStop}%
\bibitem [{\citenamefont {Cui}\ \emph {et~al.}(2025)\citenamefont {Cui},
  \citenamefont {Yang}, \citenamefont {T{\"o}lle}, \citenamefont {Ye},
  \citenamefont {Yuan}, \citenamefont {Zhai}, \citenamefont {Park},
  \citenamefont {Kim}, \citenamefont {Zhang}, \citenamefont {Lin} \emph
  {et~al.}}]{cui2025ab}%
  \BibitemOpen
  \bibfield  {author} {\bibinfo {author} {\bibfnamefont {Z.-H.}\ \bibnamefont
  {Cui}}, \bibinfo {author} {\bibfnamefont {J.}~\bibnamefont {Yang}}, \bibinfo
  {author} {\bibfnamefont {J.}~\bibnamefont {T{\"o}lle}}, \bibinfo {author}
  {\bibfnamefont {H.-Z.}\ \bibnamefont {Ye}}, \bibinfo {author} {\bibfnamefont
  {S.}~\bibnamefont {Yuan}}, \bibinfo {author} {\bibfnamefont {H.}~\bibnamefont
  {Zhai}}, \bibinfo {author} {\bibfnamefont {G.}~\bibnamefont {Park}}, \bibinfo
  {author} {\bibfnamefont {R.}~\bibnamefont {Kim}}, \bibinfo {author}
  {\bibfnamefont {X.}~\bibnamefont {Zhang}}, \bibinfo {author} {\bibfnamefont
  {L.}~\bibnamefont {Lin}},  \emph {et~al.},\ }\href
  {https://doi.org/10.1038/s41467-025-56883-x} {\bibfield  {journal} {\bibinfo
  {journal} {Nat. Commun.}\ }\textbf {\bibinfo {volume} {16}},\ \bibinfo
  {pages} {1845} (\bibinfo {year} {2025})}\BibitemShut {NoStop}%
\bibitem [{\citenamefont {Bulik}\ \emph
  {et~al.}(2014{\natexlab{a}})\citenamefont {Bulik}, \citenamefont {Scuseria},\
  and\ \citenamefont {Dukelsky}}]{bulik2014density}%
  \BibitemOpen
  \bibfield  {author} {\bibinfo {author} {\bibfnamefont {I.~W.}\ \bibnamefont
  {Bulik}}, \bibinfo {author} {\bibfnamefont {G.~E.}\ \bibnamefont {Scuseria}},
  \ and\ \bibinfo {author} {\bibfnamefont {J.}~\bibnamefont {Dukelsky}},\
  }\href {\doibase 10.1103/PhysRevB.89.035140} {\bibfield  {journal} {\bibinfo
  {journal} {Phys. Rev. B}\ }\textbf {\bibinfo {volume} {89}},\ \bibinfo
  {pages} {035140} (\bibinfo {year} {2014}{\natexlab{a}})}\BibitemShut
  {NoStop}%
\bibitem [{\citenamefont {Bulik}\ \emph
  {et~al.}(2014{\natexlab{b}})\citenamefont {Bulik}, \citenamefont {Chen},\
  and\ \citenamefont {Scuseria}}]{bulik2014electron}%
  \BibitemOpen
  \bibfield  {author} {\bibinfo {author} {\bibfnamefont {I.~W.}\ \bibnamefont
  {Bulik}}, \bibinfo {author} {\bibfnamefont {W.}~\bibnamefont {Chen}}, \ and\
  \bibinfo {author} {\bibfnamefont {G.~E.}\ \bibnamefont {Scuseria}},\ }\href
  {https://doi.org/10.1063/1.4891861} {\bibfield  {journal} {\bibinfo
  {journal} {J. Chem. Phys.}\ }\textbf {\bibinfo {volume} {141}},\ \bibinfo
  {pages} {054113} (\bibinfo {year} {2014}{\natexlab{b}})}\BibitemShut
  {NoStop}%
\bibitem [{\citenamefont {Fulde}\ and\ \citenamefont
  {Stoll}(2017)}]{fulde2017dealing}%
  \BibitemOpen
  \bibfield  {author} {\bibinfo {author} {\bibfnamefont {P.}~\bibnamefont
  {Fulde}}\ and\ \bibinfo {author} {\bibfnamefont {H.}~\bibnamefont {Stoll}},\
  }\href {https://doi.org/10.1063/1.4983207} {\bibfield  {journal} {\bibinfo
  {journal} {J. Chem. Phys.}\ }\textbf {\bibinfo {volume} {146}},\ \bibinfo
  {pages} {194107} (\bibinfo {year} {2017})}\BibitemShut {NoStop}%
\bibitem [{\citenamefont {Plat}\ and\ \citenamefont
  {Hotta}(2020)}]{plat2020entanglement}%
  \BibitemOpen
  \bibfield  {author} {\bibinfo {author} {\bibfnamefont {X.}~\bibnamefont
  {Plat}}\ and\ \bibinfo {author} {\bibfnamefont {C.}~\bibnamefont {Hotta}},\
  }\href {https://doi.org/10.1103/PhysRevB.102.140410} {\bibfield  {journal}
  {\bibinfo  {journal} {Phys. Rev. B}\ }\textbf {\bibinfo {volume} {102}},\
  \bibinfo {pages} {140410} (\bibinfo {year} {2020})}\BibitemShut {NoStop}%
\bibitem [{\citenamefont {Makhlouf}\ \emph {et~al.}(2025)\citenamefont
  {Makhlouf}, \citenamefont {Senjean},\ and\ \citenamefont {Fromager}}]{LPFET}%
  \BibitemOpen
  \bibfield  {author} {\bibinfo {author} {\bibfnamefont {W.}~\bibnamefont
  {Makhlouf}}, \bibinfo {author} {\bibfnamefont {B.}~\bibnamefont {Senjean}}, \
  and\ \bibinfo {author} {\bibfnamefont {E.}~\bibnamefont {Fromager}},\ }\href
  {\doibase 10.1021/acs.jctc.5c01256} {\bibfield  {journal} {\bibinfo
  {journal} {Journal of Chemical Theory and Computation}\ }\textbf {\bibinfo
  {volume} {21}},\ \bibinfo {pages} {10293} (\bibinfo {year}
  {2025})}\BibitemShut {NoStop}%
\bibitem [{\citenamefont {Mordovina}\ \emph {et~al.}(2019)\citenamefont
  {Mordovina}, \citenamefont {Reinhard}, \citenamefont {Theophilou},
  \citenamefont {Appel},\ and\ \citenamefont {Rubio}}]{mordovina2019self}%
  \BibitemOpen
  \bibfield  {author} {\bibinfo {author} {\bibfnamefont {U.}~\bibnamefont
  {Mordovina}}, \bibinfo {author} {\bibfnamefont {T.~E.}\ \bibnamefont
  {Reinhard}}, \bibinfo {author} {\bibfnamefont {I.}~\bibnamefont
  {Theophilou}}, \bibinfo {author} {\bibfnamefont {H.}~\bibnamefont {Appel}}, \
  and\ \bibinfo {author} {\bibfnamefont {A.}~\bibnamefont {Rubio}},\ }\href
  {\doibase 10.1021/acs.jctc.9b00063} {\bibfield  {journal} {\bibinfo
  {journal} {J. Chem. Theory Comput.}\ }\textbf {\bibinfo {volume} {15}},\
  \bibinfo {pages} {5209} (\bibinfo {year} {2019})}\BibitemShut {NoStop}%
\bibitem [{\citenamefont {Sekaran}\ \emph {et~al.}(2022)\citenamefont
  {Sekaran}, \citenamefont {Saubanère},\ and\ \citenamefont
  {Fromager}}]{sekaran2022local}%
  \BibitemOpen
  \bibfield  {author} {\bibinfo {author} {\bibfnamefont {S.}~\bibnamefont
  {Sekaran}}, \bibinfo {author} {\bibfnamefont {M.}~\bibnamefont {Saubanère}},
  \ and\ \bibinfo {author} {\bibfnamefont {E.}~\bibnamefont {Fromager}},\
  }\href {\doibase 10.3390/computation10030045} {\bibfield  {journal} {\bibinfo
   {journal} {Computation}\ }\textbf {\bibinfo {volume} {10}},\ \bibinfo
  {pages} {45} (\bibinfo {year} {2022})}\BibitemShut {NoStop}%
\bibitem [{\citenamefont {Sekaran}\ \emph {et~al.}(2023)\citenamefont
  {Sekaran}, \citenamefont {Bindech},\ and\ \citenamefont
  {Fromager}}]{sekaran2023unified}%
  \BibitemOpen
  \bibfield  {author} {\bibinfo {author} {\bibfnamefont {S.}~\bibnamefont
  {Sekaran}}, \bibinfo {author} {\bibfnamefont {O.}~\bibnamefont {Bindech}}, \
  and\ \bibinfo {author} {\bibfnamefont {E.}~\bibnamefont {Fromager}},\ }\href
  {\doibase 10.1063/5.0157746} {\bibfield  {journal} {\bibinfo  {journal} {J.
  Chem. Phys.}\ }\textbf {\bibinfo {volume} {159}},\ \bibinfo {pages} {034107}
  (\bibinfo {year} {2023})}\BibitemShut {NoStop}%
\bibitem [{\citenamefont {Sekaran}\ \emph {et~al.}(2021)\citenamefont
  {Sekaran}, \citenamefont {Tsuchiizu}, \citenamefont {Sauban\`ere},\ and\
  \citenamefont {Fromager}}]{sekaran2021householder}%
  \BibitemOpen
  \bibfield  {author} {\bibinfo {author} {\bibfnamefont {S.}~\bibnamefont
  {Sekaran}}, \bibinfo {author} {\bibfnamefont {M.}~\bibnamefont {Tsuchiizu}},
  \bibinfo {author} {\bibfnamefont {M.}~\bibnamefont {Sauban\`ere}}, \ and\
  \bibinfo {author} {\bibfnamefont {E.}~\bibnamefont {Fromager}},\ }\href
  {\doibase 10.1103/PhysRevB.104.035121} {\bibfield  {journal} {\bibinfo
  {journal} {Phys. Rev. B}\ }\textbf {\bibinfo {volume} {104}},\ \bibinfo
  {pages} {035121} (\bibinfo {year} {2021})}\BibitemShut {NoStop}%
\bibitem [{\citenamefont {Seidl}\ \emph {et~al.}(1996)\citenamefont {Seidl},
  \citenamefont {G\"orling}, \citenamefont {Vogl}, \citenamefont {Majewski},\
  and\ \citenamefont {Levy}}]{seidl1996generalized}%
  \BibitemOpen
  \bibfield  {author} {\bibinfo {author} {\bibfnamefont {A.}~\bibnamefont
  {Seidl}}, \bibinfo {author} {\bibfnamefont {A.}~\bibnamefont {G\"orling}},
  \bibinfo {author} {\bibfnamefont {P.}~\bibnamefont {Vogl}}, \bibinfo {author}
  {\bibfnamefont {J.~A.}\ \bibnamefont {Majewski}}, \ and\ \bibinfo {author}
  {\bibfnamefont {M.}~\bibnamefont {Levy}},\ }\href {\doibase
  10.1103/PhysRevB.53.3764} {\bibfield  {journal} {\bibinfo  {journal} {Phys.
  Rev. B}\ }\textbf {\bibinfo {volume} {53}},\ \bibinfo {pages} {3764}
  (\bibinfo {year} {1996})}\BibitemShut {NoStop}%
\bibitem [{\citenamefont {Löwdin}(1950)}]{lowdin1950non}%
  \BibitemOpen
  \bibfield  {author} {\bibinfo {author} {\bibfnamefont {P.}~\bibnamefont
  {Löwdin}},\ }\href {\doibase 10.1063/1.1747632} {\bibfield  {journal}
  {\bibinfo  {journal} {J. Chem. Phys.}\ }\textbf {\bibinfo {volume} {18}},\
  \bibinfo {pages} {365} (\bibinfo {year} {1950})}\BibitemShut {NoStop}%
\bibitem [{\citenamefont {Yalouz}\ \emph
  {et~al.}(2022{\natexlab{a}})\citenamefont {Yalouz}, \citenamefont {Gullin},\
  and\ \citenamefont {Sekaran}}]{yalouz2022quantnbody}%
  \BibitemOpen
  \bibfield  {author} {\bibinfo {author} {\bibfnamefont {S.}~\bibnamefont
  {Yalouz}}, \bibinfo {author} {\bibfnamefont {M.~R.}\ \bibnamefont {Gullin}},
  \ and\ \bibinfo {author} {\bibfnamefont {S.}~\bibnamefont {Sekaran}},\ }\href
  {https://doi.org/10.21105/joss.04759} {\bibfield  {journal} {\bibinfo
  {journal} {J. Open Source Softw.}\ }\textbf {\bibinfo {volume} {7}},\
  \bibinfo {pages} {4759} (\bibinfo {year} {2022}{\natexlab{a}})}\BibitemShut
  {NoStop}%
\bibitem [{\citenamefont {Smith}\ \emph {et~al.}(2020)\citenamefont {Smith},
  \citenamefont {Burns}, \citenamefont {Simmonett}, \citenamefont {Parrish},
  \citenamefont {Schieber}, \citenamefont {Galvelis}, \citenamefont {Kraus},
  \citenamefont {Kruse}, \citenamefont {Di~Remigio}, \citenamefont {Alenaizan}
  \emph {et~al.}}]{smith2020psi4}%
  \BibitemOpen
  \bibfield  {author} {\bibinfo {author} {\bibfnamefont {D.~G.}\ \bibnamefont
  {Smith}}, \bibinfo {author} {\bibfnamefont {L.~A.}\ \bibnamefont {Burns}},
  \bibinfo {author} {\bibfnamefont {A.~C.}\ \bibnamefont {Simmonett}}, \bibinfo
  {author} {\bibfnamefont {R.~M.}\ \bibnamefont {Parrish}}, \bibinfo {author}
  {\bibfnamefont {M.~C.}\ \bibnamefont {Schieber}}, \bibinfo {author}
  {\bibfnamefont {R.}~\bibnamefont {Galvelis}}, \bibinfo {author}
  {\bibfnamefont {P.}~\bibnamefont {Kraus}}, \bibinfo {author} {\bibfnamefont
  {H.}~\bibnamefont {Kruse}}, \bibinfo {author} {\bibfnamefont
  {R.}~\bibnamefont {Di~Remigio}}, \bibinfo {author} {\bibfnamefont
  {A.}~\bibnamefont {Alenaizan}},  \emph {et~al.},\ }\href
  {https://doi.org/10.1063/5.0006002} {\bibfield  {journal} {\bibinfo
  {journal} {The Journal of chemical physics}\ }\textbf {\bibinfo {volume}
  {152}} (\bibinfo {year} {2020})}\BibitemShut {NoStop}%
\bibitem [{\citenamefont {Yalouz}\ \emph
  {et~al.}(2022{\natexlab{b}})\citenamefont {Yalouz}, \citenamefont {Sekaran},
  \citenamefont {Fromager},\ and\ \citenamefont
  {Saubanère}}]{yalouz2022quantum}%
  \BibitemOpen
  \bibfield  {author} {\bibinfo {author} {\bibfnamefont {S.}~\bibnamefont
  {Yalouz}}, \bibinfo {author} {\bibfnamefont {S.}~\bibnamefont {Sekaran}},
  \bibinfo {author} {\bibfnamefont {E.}~\bibnamefont {Fromager}}, \ and\
  \bibinfo {author} {\bibfnamefont {M.}~\bibnamefont {Saubanère}},\ }\href
  {\doibase 10.1063/5.0125683} {\bibfield  {journal} {\bibinfo  {journal} {J.
  Chem. Phys.}\ }\textbf {\bibinfo {volume} {157}} (\bibinfo {year}
  {2022}{\natexlab{b}}),\ 10.1063/5.0125683}\BibitemShut {NoStop}%
\bibitem [{\citenamefont {Nusspickel}\ \emph {et~al.}(2023)\citenamefont
  {Nusspickel}, \citenamefont {Ibrahim},\ and\ \citenamefont
  {Booth}}]{nusspickel2023effective}%
  \BibitemOpen
  \bibfield  {author} {\bibinfo {author} {\bibfnamefont {M.}~\bibnamefont
  {Nusspickel}}, \bibinfo {author} {\bibfnamefont {B.}~\bibnamefont {Ibrahim}},
  \ and\ \bibinfo {author} {\bibfnamefont {G.~H.}\ \bibnamefont {Booth}},\
  }\href {\doibase 10.1021/acs.jctc.2c01063} {\bibfield  {journal} {\bibinfo
  {journal} {J. Chem. Theory Comput.}\ }\textbf {\bibinfo {volume} {19}},\
  \bibinfo {pages} {2769} (\bibinfo {year} {2023})}\BibitemShut {NoStop}%
\bibitem [{\citenamefont {Wesolowski}(2025)}]{Wesolowski25_Density}%
  \BibitemOpen
  \bibfield  {author} {\bibinfo {author} {\bibfnamefont {T.~A.}\ \bibnamefont
  {Wesolowski}},\ }\href {\doibase 10.1063/5.0279936} {\bibfield  {journal}
  {\bibinfo  {journal} {The Journal of Chemical Physics}\ }\textbf {\bibinfo
  {volume} {163}},\ \bibinfo {pages} {164112} (\bibinfo {year}
  {2025})}\BibitemShut {NoStop}%
\bibitem [{\citenamefont {Lanat\`a}(2023)}]{lanata2023derivation}%
  \BibitemOpen
  \bibfield  {author} {\bibinfo {author} {\bibfnamefont {N.}~\bibnamefont
  {Lanat\`a}},\ }\href {\doibase 10.1103/PhysRevB.108.235112} {\bibfield
  {journal} {\bibinfo  {journal} {Phys. Rev. B}\ }\textbf {\bibinfo {volume}
  {108}},\ \bibinfo {pages} {235112} (\bibinfo {year} {2023})}\BibitemShut
  {NoStop}%
\bibitem [{\citenamefont {Giuli}\ \emph {et~al.}(2025)\citenamefont {Giuli},
  \citenamefont {Hasan}, \citenamefont {Kloss}, \citenamefont {Frank},
  \citenamefont {Lee}, \citenamefont {Gingras}, \citenamefont {Yao},\ and\
  \citenamefont {Lanatà}}]{giuli2025linearfoundationmodelquantum}%
  \BibitemOpen
  \bibfield  {author} {\bibinfo {author} {\bibfnamefont {S.}~\bibnamefont
  {Giuli}}, \bibinfo {author} {\bibfnamefont {H.}~\bibnamefont {Hasan}},
  \bibinfo {author} {\bibfnamefont {B.}~\bibnamefont {Kloss}}, \bibinfo
  {author} {\bibfnamefont {M.~S.}\ \bibnamefont {Frank}}, \bibinfo {author}
  {\bibfnamefont {T.-H.}\ \bibnamefont {Lee}}, \bibinfo {author} {\bibfnamefont
  {O.}~\bibnamefont {Gingras}}, \bibinfo {author} {\bibfnamefont {Y.-X.}\
  \bibnamefont {Yao}}, \ and\ \bibinfo {author} {\bibfnamefont
  {N.}~\bibnamefont {Lanatà}},\ }\href {https://arxiv.org/abs/2512.21666}
  {\enquote {\bibinfo {title} {Linear foundation model for quantum embedding:
  Data-driven compression of the ghost gutzwiller variational space},}\ }
  (\bibinfo {year} {2025}),\ \Eprint {http://arxiv.org/abs/2512.21666}
  {arXiv:2512.21666 [cond-mat.str-el]} \BibitemShut {NoStop}%
\end{thebibliography}

\newcommand{\Aa}[0]{Aa}
%


\end{document}